\documentclass[twocolumn,aps,superscriptaddress,preprintnumbers]{revtex4-2}

\usepackage{amsmath,amssymb}
\usepackage{graphicx}
\usepackage{dcolumn} 
\usepackage{bm} 
\usepackage{soul}
\usepackage{multirow}
\usepackage{setspace}
\usepackage[mathlines]{lineno}
\usepackage{textcomp}
\usepackage{mathcomp}
\usepackage{spverbatim}
\usepackage{float}
\usepackage{wasysym}
\usepackage{array}
\usepackage{subfigure}
\usepackage{physics}
\usepackage{xcolor}

\usepackage{hyperref}
\hypersetup{
    colorlinks=true,       
    linkcolor=blue,        
    citecolor=blue,        
    filecolor=blue,        
    urlcolor=blue,         
    runcolor=blue
}

\usepackage{enumitem,amssymb}

\begin{document}

\title{Bath-induced deviations from Gibbs statistics for strongly interacting oscillators}

\author{Felipe Recabal}
\affiliation{Department of Physics, Universidad de Santiago de Chile, Av. Victor Jara 3493, Santiago, Chile.}
\author{Adrian E. Rubio Lopez}
\affiliation{Department of Physics, Universidad de Santiago de Chile, Av. Victor Jara 3493, Santiago, Chile.}
\affiliation{Millennium Institute for Research in Optics, Concepci\'on, Chile.}
\author{Johannes Schachenmayer}
\affiliation{CESQ/ISIS (UMR 7006), CNRS and Université de Strasbourg, 67000 Strasbourg, France}
\author{Felipe Herrera}
\affiliation{Department of Physics, Universidad de Santiago de Chile, Av. Victor Jara 3493, Santiago, Chile.}
\affiliation{Millennium Institute for Research in Optics, Concepci\'on, Chile.}

\date{\today}

\begin{abstract}
The Redfield quantum master equation is widely used to study the dynamics of interacting subsystems that are weakly coupled to baths. Redfield dynamics under secular approximation preserves positivity of the reduced density operator and thermalizes the system into a Gibbs state at equilibrium. Long-time effects arising from non-secular terms are often neglected, but depending on the system spectrum and relative bath couplings, non-secular contributions are shown here to drive the system into a non-Gibbs state. For two strongly interacting quantum oscillators with independent baths at equal temperature, we analyze the microscopic origin of the deviations from Gibbs statistics. Provided that the oscillators are unequally damped by their baths, we show that steady state occupation numbers can significantly deviate from a Boltzmann distribution due to an excitation flux driven by bath-induced coherences between nearly-degenerate oscillator levels. Conditions for the recovery of thermal Gibbs statistics are discussed and experimental signatures suggested. 
\end{abstract}

\maketitle

\section{Introduction}

Master equations are used to model open quantum systems, describing the effects of the environment on various features of a target quantum system, such as coherence and entanglement \cite{eastham2016bath,Sousa,purkayastha2020tunable}, which are crucial resources in areas such as quantum metrology \cite{pires2018coherence,joo2011quantum} and quantum computing \cite{albash2015decoherence}. Quantum master equations are often constructed in Lindblad form, providing a Markovian evolution that ensures positive system probabilities and provides direct description of system relaxation and decoherence induced by the environment \cite{albash2012quantum,trushechkin2022open,davidovic2020completely}, being applied across a variety of scenarios \cite{purkayastha2020tunable,ahn2023modification,roy2011influence,palmieri2009lindblad,Herrera2020,scali2021local,recabal2026,orgiu2015}. For systems composed of multiple weakly interacting subsystems, local-Lindblad master equation describes the influence of each bath on a given subsystem as if they were isolated from each other. However, for strongly interacting subsystems, this approach can lead to incorrect dynamics and steady states with thermodynamic inconsistencies \cite{Joshi,scali2021local,cattaneo2019local}, such as violations of the second law of thermodynamics \cite{levy2014local}. A general approach to describe strongly interacting subsystems is provided by the Redfield master equation, which microscopically incorporates the subsystem couplings. The Redfield and local-Lindblad approaches yield different predictions in scenarios of strongly interacting subsystems \cite{cresser1992thermal,Sousa,hartmann2020accuracy,ahn2023modification,Beaudoin2011}, as observed in experiments such as circuit \cite{baust2016ultrastrong,wang2020photon,wallraff2004,mergenthaler2017,delrio2026}  and cavity \cite{takahashi2020strong,ahn2023modification,triana2022,chikkarddy2016} quantum electrodynamics.

Redfield master equations do not guarantee positive probabilities, an issue that is commonly addressed by applying a secular approximation that neglects population–coherence couplings in the energy basis of the system. This approximation enforces strictly positive system probabilities, as well as relaxation of the system into a Gibbs state at the bath temperature \cite{dann2021open,trushechkin2022open,Sousa}. However, non-secular terms are in principle non-zero and could give rise to population–coherence couplings that have been shown to be relevant for long-time and steady state predictions \cite{scali2021local,eastham2016bath,cattaneo2019local,cresser2017coarse,tscherbul2015partial,mori2008dynamics,geva2000second,albash2012quantum,de2010effects,farina2019open}, for example in light-harvesting systems \cite{jeske2015bloch,palmieri2009lindblad}.  
In scenarios in which the secular approximation breaks down, such as in multiple interacting subsystems, a way to preserve physical relevant non-secular terms and avoid negative probabilities is by coarse-graining or partial secularizing the Redfield dynamics \cite{tscherbul2015partial,schaller2008preservation,cattaneo2019local,davidovic2020completely,farina2019open,hartmann2020accuracy,Mozgunov2020}, leading to open questions about the predicted system steady states and their thermodynamic consistencies \cite{becker2022canonically,cattaneo2019local,potts2021thermodynamically}. Efforts to address these questions include the use of the mean force Gibbs state, which is a partial trace over the entire system-bath Hilbert space open assumed to have Gibbs quantum statistics. The mean force Gibbs state  in general deviates from Gibbs statistics as a consequence of system-bath interaction \cite{cresser2021weak,trushechkin2022open,becker2022canonically,Brenes2024}. 

In this context, we study the steady state properties of two strongly interacting quantum harmonic oscillators, each of them being weakly coupled to an independent thermal bath. The baths for both oscillators have equal temperature (thermal equilibrium), but the oscillators can couple to them with different strengths (spectral densities). We show that the non-secular terms of a coarse-grained Redfield quantum master equation lead to equilibration of the coupled system to a non-Gibbs state, provided that the oscillators unequally couple to their baths. This deviation from canonical Gibbs statistics manifests a stationary deviation of the occupation numbers from the thermal Boltzmann distribution, sustained by a steady excitation flux established between the baths only when the oscillators are resonantly coupled. 

The rest of the paper is organized as follows: In Sec. \ref{sec:methods} we describe our open quantum system model, and derive the system and bath dynamics using a partial secular Redfield approach. In Sec. \ref{sec:steady_state}  we study the steady state predictions of our model, giving parameter conditions and regimens where steady system and bath features of the non-Gibbs state can be observed. Conclusions are given in Sec. \ref{sec:conclusion}.

\section{Open Quantum System Model}\label{sec:methods}

We consider an open quantum system described by the Hamiltonian $\hat{\mathcal{H}} =\hat{\mathcal{H}}_S + \hat{\mathcal{H}}_B + \hat{\mathcal{H}}_{SB}$ (units of $\hbar = 1$ and $k_B=1$ are used throughout this manuscript). The system, described by the system density operators $\hat{\rho}_S$ and the Hamiltonian $\hat{\mathcal{H}}_S$, is weakly coupled to thermal equilibrium baths at the common temperature $T=1/\beta$, described by the bath density operator $\hat{\rho}_B \sim \exp (-\beta \hat{\mathcal{H}}_B)$ and the Hamiltonian $\hat{\mathcal{H}}_B$. In the Born-Markov approximation, the system dynamics is given by the Redfield equation, which incorporates the effect of the system-bath interaction, described by the Hamiltonian $\hat{\mathcal{H}}_{SB}$, as a second order perturbation.

\subsection{Hamiltonian and Spectral Densities}\label{sec:Hamiltonians}
Specifically, our system is composed of two quantum harmonic oscillators, A and B, with frequencies $\omega_a$ and $\omega_b$, respectively. The oscillators interact via the coupling strength $g$, leading to the system Hamiltonian
\begin{equation}\label{eq:system_Hamiltonian}
\hat{\mathcal{H}}_S = \omega_a \hat{a}^\dagger \hat{a}+\omega_b \hat{b}^\dagger \hat{b} + g(\hat{a} + \hat{a}^\dagger  )(\hat{b} + \hat{b}^\dagger ).
\end{equation}
Operators $\hat{a}$ and $\hat{b}$ annihilate bosons of the modes $\omega_a$ and $\omega_b$, respectively. We consider interacting oscillators in the strong-coupling regime, i.e., $g\ll \{  \omega_a, \omega_b\}$, where counter-rotating terms $\hat{a}\hat{b}$ and $\hat{a}^\dagger \hat{b}^\dagger$ in Eq. (\ref{eq:system_Hamiltonian}) are ignored \cite{Sousa,Joshi,Beaudoin2011}. However, for a general analysis, part of the derivations in this work, we will include the counter-rotating terms.

The oscillators A and B are in contact with their independent baths, which are thermal equilibrium reservoirs of oscillators, described by the Hamiltonian $\hat{\mathcal{H}}_B =  \sum_i \omega_i \hat{a}_i^\dagger\hat{a}_i + \sum_j \omega_j \hat{b}_j^\dagger\hat{b}_j$. Operators $\hat{a}_i$  and $\hat{b}_j$ are ladder operators for the bath frequency modes $\omega_i$ and $\omega_j$, in the respective two baths. The interaction between oscillators and their baths is described as
\begin{eqnarray}\label{eq:interaction_Hamiltonian}
\hat{\mathcal{H}}_{SB} =&   (\hat{a} + \hat{a}^\dagger)\otimes \sum_i \lambda_i^{(a)} \left(\hat{a}_i + \hat{a}_i^\dagger  \right) \nonumber\\
+& (\hat{b} + \hat{b}^\dagger)\otimes \sum_j \lambda_j^{(b)} \left(\hat{b}_j + \hat{b}_j^\dagger  \right),
\end{eqnarray}
where $\lambda_i^{(a)}$ and $\lambda_j^{(b)}$ are the coupling strengths of oscillators A and B with their respective baths. We model system-bath coupling strengths of oscillator A and B by the Ohmic spectral densities
\begin{eqnarray}
    \mathcal{J}_a(\omega) = \sum_i |\lambda_i^{(a)} |^2 \delta(\omega - \omega_i)= \eta_a\omega e^{-\omega/\omega_c}, \label{eq:Ohmic_a} \\
    \mathcal{J}_b(\omega) = \sum_j |\lambda_j^{(b)} |^2\delta(\omega - \omega_j) = \eta_b\omega e^{-\omega/\omega_c}, \label{eq:Ohmic_b}
\end{eqnarray}
respectively. In Eq. (\ref{eq:Ohmic_a}) and (\ref{eq:Ohmic_b}), $\eta_a$ and $\eta_b$ are the system-bath coupling strengths and $\omega_{c}$ is a common cutoff frequency. For resonant non-interacting oscillators ($\omega_a = \omega_b$ and $g=0$), the baths induce damping on the dynamics of oscillators A and B with associated damping rates $\gamma_a = 2\pi \mathcal{J}_a(\omega_b)$ and $\gamma_b = 2\pi \mathcal{J}_b(\omega_b)$, respectively. In our model, it implies that the strengths $\eta_a \propto \gamma_a$ and $\eta_b \propto \gamma_b$ are proportional to the damping rates, reducing Eq. (\ref{eq:Ohmic_a}) and (\ref{eq:Ohmic_b}) to the structure
\begin{equation}\label{eq:dimensionless_spectral_density}
    \mathcal{J}(\omega) = \frac{\mathcal{J}_\alpha (\omega)}{\gamma_\alpha} = \frac{1}{2\pi} \frac{\omega}{\omega_b}e^{-(\omega - \omega_b)/\omega_c},
\end{equation}
for $\alpha=\{a,b \}$ referring to the oscillator A or B. In Eq. (\ref{eq:dimensionless_spectral_density}), the dimensionless function $\mathcal{J}(\omega)$ contains the Ohmic dependence of both baths.

In our model, oscillators could differ in frequencies, associated to the frequency detuning $\Delta = \omega_a-\omega_b$, and damping rates, $\gamma_a$ and $\gamma_b$. The regime of weak system-bath coupling and strong oscillator coupling is reduced to the condition $\{ \gamma_a,\gamma_b \} \ll g \ll \{ \omega_a , \omega_b \}$, which is fulfilled in the rest of the analysis. This parameter regime could be applicable to experimental platforms where strong-light matter coupling is observed, such as superconductor resonators \cite{delrio2026,takahashi2020strong,mergenthaler2017,wang2020,baust2016ultrastrong,wallraff2004}, plasmonic nanocavities \cite{chikkarddy2016,triana2022,yu2023ostensible}, or Fabry-Perot cavities \cite{long2015,erwin2019,ahn2023modification}.

\subsection{Coarse-Grained Redfield Equation}
The derivation of Redfield system dynamics requires expressing the system Hamiltonian in diagonal form. Reducing the interacting oscillators in Eq. (\ref{eq:system_Hamiltonian}) to two normal modes, with frequencies $\Omega_1$ and $\Omega_2$, we transform the system Hamiltonian to
\begin{equation}\label{eq:diagonal_Hamiltonian}
\hat{\mathcal{H}}_S =  \Omega_{1} \hat{c}_1^\dagger \hat{c}_1 + \Omega_{2}\hat{c}_2^\dagger \hat{c}_2.
\end{equation}
Here, normal modes are described by the mode ladder operators $\hat{c}_1$ and $\hat{c}_2$, which are linear combinations of $\hat{a}$ and $\hat{b}$ \cite{xiao2009theory}. In the strong coupling regime, i.e., neglecting counter-rotating terms, the transformations are $\hat{a} =  \cos(A)\hat{c}_1  - \sin(A)\hat{c}_2$ and $\hat{b} =  \cos(A) \hat{c}_2 + \sin(A)\hat{c}_1$, with $A = \arctan(2g/\Delta)/2$.
The associated normal mode frequencies are
\begin{eqnarray}
    \Omega_{1} = \omega_a + \frac{1}{2}( \sqrt{4g^2 + \Delta^2} - \Delta  ), \label{eq:omega_1}\\ 
    \Omega_{2} = \omega_b - \frac{1}{2}( \sqrt{4g^2 + \Delta^2} - \Delta ),\label{eq:omega_2}
\end{eqnarray}
and depend on the frequency detuning $\Delta$, featuring a (Rabi) splitting $\Omega_1 - \Omega_2= 2g$ in resonance (for $\Delta=0$).

The impact of the thermal baths on the system dynamics can be described in terms of the dimensionless function
\begin{equation}\label{eq:bath_correlations}
\begin{split}
    \mathcal{C}(\omega) = \frac{1}{2}\Gamma (\omega) +   i \mathcal{S} (\omega),
\end{split}
\end{equation}
where the real part of Eq. (\ref{eq:bath_correlations}),
\begin{equation}\label{eq:def_rates}
    \Gamma(\omega) = \begin{cases}
2\pi \mathcal{J}(|\omega|)\Big[  1+ n^{\rm th}(|\omega|) \Big]  &,\hspace{0.2cm}\omega > 0\\
2\pi \mathcal{J}(|\omega|) n^{\rm th}(|\omega|)  &, \hspace{0.2cm}\omega < 0
\end{cases}
\end{equation}
is written in terms of $\mathcal{J}(\omega)$ from Eq. (\ref{eq:dimensionless_spectral_density}) and the thermal occupation $n^{\rm th}(\omega)=[\exp(\beta \omega)-1]^{-1}$ at frequency $\omega$. Meanwhile, the imaginary part of Eq. (\ref{eq:bath_correlations}),
\begin{eqnarray}\label{eq:def_lamb_shifts}
    \mathcal{S} (\omega) = \mathcal{P} \int_0^\infty d\omega' \mathcal{J} (\omega') \Bigg\{ \frac{n^{\rm th}(\omega') }{\omega' +\omega } - \frac{\left[ 1+n^{\rm th}(\omega')\right] }{\omega' - \omega}\Bigg\},
\end{eqnarray}
is defined by a Cauchy principal value, denoted by $\mathcal{P}$.

In terms of system normal modes in Eq. (\ref{eq:diagonal_Hamiltonian}) and the bath function in Eq. (\ref{eq:bath_correlations}), we derive the system dynamics in the Redfield approach (for a full derivation, see Appendix \ref{appendix:Redfield_QME}). The resulting Redfield master equation contains time-independent (secular) and time-dependent (non-secular) terms, with non-secular terms oscillating at frequencies $ \{  2 \Omega_1, 2\Omega_2, \Omega_1 + \Omega_2,\Omega_1 - \Omega_2 \}$. In order to avoid possible negative probabilities in the system density operator, we employed a partial secularization of the Redfield dynamics. Specifically, when the strongly interacting oscillators are nearly resonant (condition $|\Delta| < g$ in Eq. (\ref{eq:omega_1}) and (\ref{eq:omega_2})), the frequencies $\Omega_1 \approx \omega_b +g$ and $\Omega_2 \approx \omega_b -g$ justify the preservation of the slower oscillating non-secular terms with frequency $\Omega_1 - \Omega_2 = 2g$, while the rest are ignored (details in Appendix \ref{appendix:partial_secular}). The resulting partial secular Redfield master equation \cite{cattaneo2019local,cresser2017coarse,hartmann2020accuracy}
\begin{widetext}
    \begin{equation}\label{eq:CG_Redfield}
    \begin{split}
    \frac{\rm d}{\rm dt}\hat{\rho}_S =  -& i\overline{\Omega}_1[\hat{c}_1^\dagger \hat{c}_1,\hat{\rho}_S ] + \Gamma_1(\Omega_1) \mathcal{L}_{ \hat{c}_1} [\hat{\rho}_S]+ \Gamma_1(-\Omega_1)  \mathcal{L}_{ \hat{c}_1^\dagger} [\hat{\rho}_S] - i \overline{\Omega}_2[\hat{c}_2^\dagger \hat{c}_2,\hat{\rho}_S ] + \Gamma_2(\Omega_2)\mathcal{L}_{ \hat{c}_2} [\hat{\rho}_S] + \Gamma_2(-\Omega_2) \mathcal{L}_{ \hat{c}_2^\dagger} [\hat{\rho}_S] \\
    -& i\overline{g}[\hat{c}_1 \hat{c}_2^\dagger,\hat{\rho}_S ] +  \Gamma_{12} \mathcal{L}_{\hat{c}_2^\dagger ,\hat{c}_1}[\hat{\rho}_S]  + \Gamma_{21}^* \mathcal{L}_{\hat{c}_1,\hat{c}_2^\dagger}[\hat{\rho}_S]  - i\overline{g}^*[\hat{c}_1 ^\dagger \hat{c}_2 ,\hat{\rho}_S ] + \Gamma_{21} \mathcal{L}_{\hat{c}_2 ,\hat{c}_1^\dagger}[\hat{\rho}_S] + \Gamma_{12}^* \mathcal{L}_{\hat{c}_1^\dagger ,\hat{c}_2}[\hat{\rho}_S],
    \end{split}
    \end{equation}
\end{widetext}
describes the system dynamics in the normal mode basis. The secular contributions in the system dynamics (first line in Eq. (\ref{eq:CG_Redfield})) describe the Lamb shifts of normal mode frequencies, according to
\begin{eqnarray}
\overline{\Omega}_1 = \Omega_1 + \left[ \gamma_a\cos^2(A) + \gamma_b  \sin^2(A) \right]\left[ \mathcal{S}(\Omega_1) + \mathcal{S}(-\Omega_1) \right], \nonumber \\
\label{eq:modified_omega_1}\\
\overline{\Omega}_2 = \Omega_2 + \left[ \gamma_b\cos^2(A)+ \gamma_a\sin^2(A) \right] \left[ \mathcal{S}(\Omega_2) + \mathcal{S}(-\Omega_2) \right] \nonumber,\\
\label{eq:modified_omega_2}
\end{eqnarray}
and excitation and de-excitation of normal modes, with associated rates
\begin{eqnarray}
\Gamma_1(\omega) = \left[ \gamma_a\cos^2(A)  + \gamma_b\sin^2(A)  \right] \Gamma( \omega),\label{eq:rate_1}\\
\Gamma_2( \omega) = \left[ \gamma_b \cos^2(A)  + \gamma_a \sin^2(A) \right] \Gamma(\omega). \label{eq:rate_2}
\end{eqnarray}
The secular dissipation is written in terms of the Lindblad super-operator
\begin{equation}\label{eq:Lindblad}
\mathcal{L}_{\hat{c}_i} [\hat{\rho}_S] = \hat{c}_i \hat{\rho}_S \hat{c}_i ^\dagger - \frac{1}{2} \Big\{ \hat{c}_i^\dagger \hat{c}_i,\hat{\rho}_S  \Big\}.
\end{equation}
The rates of the secular processes are real functions constructed as combinations of real or imaginary parts of $\mathcal{C}(\omega)$ in Eq. (\ref{eq:bath_correlations}).

The non-secular contributions in the system dynamics (second line in Eq. (\ref{eq:CG_Redfield})) describe coherent coupling between normal modes, with a coupling strength 
\begin{eqnarray}\label{eq:def_g_12}
\overline{g} &= &i\frac{g }{\sqrt{4g^2 + \Delta^2}}\left(\gamma_b - \gamma_a \right) \nonumber \\
& \times &\Big\{ \mathcal{C}(\Omega_2)^* - \mathcal{C}(\Omega_1) + \mathcal{C}(-\Omega_1)^* - \mathcal{C}(-\Omega_2) \Big\},
\end{eqnarray}
and cross dissipation between normal modes, with the associated rates 
\begin{eqnarray}
\Gamma_{12} = \frac{g }{\sqrt{4g^2 + \Delta^2}} \left(\gamma_b - \gamma_a \right)\Big\{ \mathcal{C}(\Omega_1) + \mathcal{C}(\Omega_2)^* \Big\},\label{eq:def_eta_12}\\
\Gamma_{21} = \frac{g }{\sqrt{4g^2 + \Delta^2}} \left(\gamma_b - \gamma_a \right) \Big\{ \mathcal{C}(-\Omega_1) + \mathcal{C}(-\Omega_2)^* \Big\}.\label{eq:def_eta_21}
\end{eqnarray}
The non-secular dissipation is written in terms of the super-operator
\begin{equation}\label{eq:cross_Lindblad}
\mathcal{L}_{\hat{c}_i,\hat{c}_j} [\hat{\rho}_S] = \hat{c}_j \hat{\rho}_S \hat{c}_i - \frac{1}{2} \Big\{ \hat{c}_i \hat{c}_j,\hat{\rho}_S  \Big\}.
\end{equation}
The rates of the non-secular processes are in general complex functions constructed as combinations of $\mathcal{C}(\omega)$. Although the inclusion of non-secular terms in the partial secular approximation is justified when $|\Delta| <g$, their contributions vanish for larger detuning due to the factor  $\cos(A)\sin(A)=g/\sqrt{4g^2 + \Delta^2}$ (see Eq. (\ref{eq:def_g_12}-\ref{eq:def_eta_21})). Therefore, we extend the use of the system dynamics in Eq. (\ref{eq:CG_Redfield}) for a wider range of detuning. 

We note that the dynamics of Eq. (\ref{eq:CG_Redfield}) reduces to the correct dynamics of non-interacting oscillators for $g=0$. In this case, the values $\cos(A)=1$ and $\sin(A)=0$ lead to normal mode variables $\{ \hat{c}_1,\Omega_1,\hat{c}_2,\Omega_2 \} \rightarrow \{ \hat{a},\omega_a,\hat{b},\omega_b \} $ being reduced to the individual oscillator variables, and also non-secular terms vanish. It results in a local-Lindblad master equation that contains bath-induced Lamb-shift, and excitation and de-excitation of oscillators A and B in the Lindblad form \cite{cattaneo2019local,scali2021local,trushechkin2022open,Beaudoin2011}.

\subsection{System-Bath Excitation Currents}
The dynamics of the system implies an exchange of excitations with the baths. It produces a dynamical variation of the mean number of excitations in the baths coupled to the oscillators A and B, $ \langle \hat{\mathcal{N}}_a \rangle$ and $ \langle \hat{\mathcal{N}}_b \rangle$, respectively, with $\hat{\mathcal{N}}_a = \sum_i \hat{a}_i^\dagger \hat{a}_i$ and $\hat{\mathcal{N}}_b = \sum_j \hat{b}_j^\dagger \hat{b}_j$. The partial secular Redfield approach leads to the evolution (see Appendix \ref{appendix:conservation of particles} for a full derivation)
\begin{eqnarray}
\frac{\rm d}{\rm dt} \langle \hat{\mathcal{N}}_a \rangle & = & 2 \pi\gamma_a\cos^2(A)   \mathcal{J}(\Omega_1) \left[ \langle \hat{c}_1^\dagger \hat{c}_1 \rangle - n^{\rm th}(\Omega_1) \right] \nonumber\\
&+& 2\pi \gamma_a\sin^2(A) \mathcal{J}(\Omega_2) \left[ \langle \hat{c}_2^\dagger \hat{c}_2 \rangle - n^{\rm th}(\Omega_2) \right]\nonumber \\
& - & \frac{2\gamma_a}{(\gamma_b - \gamma_a)}{\rm Re} \Big\{ \left[ \Gamma_{12} - \Gamma_{21}^*\right] \langle \hat{c}_1 \hat{c}_2^\dagger \rangle \Big\}, \label{eq:bath_particles_a}\\
\frac{\rm d}{\rm dt} \langle \hat{\mathcal{N}}_b \rangle & = & 2 \pi\gamma_b\cos^2(A) \mathcal{J}(\Omega_2) \left[ \langle \hat{c}_2^\dagger \hat{c}_2 \rangle - n^{\rm th}(\Omega_2) \right]\nonumber \\
&+& 2\pi\gamma_b \sin^2(A) \mathcal{J}(\Omega_1) \left[ \langle \hat{c}_1^\dagger \hat{c}_1 \rangle - n^{\rm th}(\Omega_1) \right] \nonumber \\
& + & \frac{2\gamma_b}{(\gamma_b - \gamma_a)}{\rm Re} \Big\{ \left[ \Gamma_{12} - \Gamma_{21}^*\right] \langle \hat{c}_1 \hat{c}_2^\dagger \rangle \Big\}, \label{eq:bath_particles_b}
\end{eqnarray}
Bath dynamics in Eq. (\ref{eq:bath_particles_a}) and (\ref{eq:bath_particles_b})  depends on the system dynamics through normal mode occupations $\langle \hat{c}_1^\dagger \hat{c}_1 \rangle$ and $\langle \hat{c}_2^\dagger \hat{c}_2 \rangle$ as the secular contributions and normal mode coherence $\langle \hat{c}_1 \hat{c}_2^\dagger \rangle$ as the non-secular contributions.

\section{Results}\label{sec:steady_state}

When ignoring the non-secular contributions in Eq. (\ref{eq:CG_Redfield}),  normal modes evolve independently. Under this secular approximation, the detailed balance condition $\Gamma (-\omega) / \Gamma(\omega) = \exp (-\beta \omega)$ (see Eq. (\ref{eq:def_rates})) relaxes the system dynamics to the Gibbs state \cite{scali2021local,Sousa}
\begin{equation}\label{eq:canonical_state}
\hat{\rho}_{G} = \frac{e^{-\beta \hat{\mathcal{H}}_S}}{{\rm Tr}_S \{ e^{-\beta \hat{\mathcal{H}}_S}\} },
\end{equation}
written in terms of the system Hamiltonian $\hat{\mathcal{H}}_S$ and the common bath temperature $T=1/\beta$. ${\rm Tr}_S$ is the trace over the system degrees of freedom. However, the Gibbs state relaxation does not incorporate possible long-time effects associated with the preserved non-secular terms.

\subsection{Non-Gibbs steady state}\label{sec:non-Gibbs state}

We identify discrepancies between the Gibbs state and the steady state of the partial secular Redfield master equation by evaluating Eq. (\ref{eq:canonical_state}) in Eq. (\ref{eq:CG_Redfield}). The derivation, detailed in the Appendix \ref{appendix:non-canonical state}, leads to the expression
\begin{eqnarray}\label{eq:non-canonical}
    \frac{\rm d}{\rm dt} \hat{\rho}_S \Bigg|_{\hat{\rho}_S= \hat{\rho}_{G}} &=& i \frac{g}{\sqrt{4g^2 + \Delta^2}}\left( \gamma_b - \gamma_a \right) \\
    &\times& \Bigg[ \frac{f(\Omega_1) }{n^{\rm th}(\Omega_2)} -  \frac{f(\Omega_2)}{n^{\rm th}(\Omega_1)}  \Bigg]\Big[  \hat{c}_2 \hat{\rho}_G \hat{c}_1^\dagger - \hat{c}_1 \hat{\rho}_G \hat{c}_2^\dagger \Big],\nonumber
\end{eqnarray}
where
\begin{eqnarray}\label{eq:def_f}
    &f (\omega) = \mathcal{S} (\omega) + e^ {\beta \omega} \mathcal{S}(-\omega).\label{eq:f}
\end{eqnarray}
The analysis of the right side of Eq. (\ref{eq:non-canonical}) establishes conditions where the system steady state deviates from the Gibbs state. Ignoring the operator contributions, the non-zero value of the right side of Eq. (\ref{eq:non-canonical}) implies that, in general, the dynamics of the strongly interacting oscillators relaxes in a non-Gibbs state. Therefore, for the rest of the work, we refer to the steady state of the system dynamics in Eq. (\ref{eq:CG_Redfield}) as the non-Gibbs state. Equation (\ref{eq:non-canonical}) is constructed as a joint effect of imaginary part of bath correlations $\mathcal{S}(\omega)$, explicitly contained in $f(\omega)$, and non-secular terms in the system dynamics. The relation of both has previously been suggested as a source of long-time modified states \cite{trushechkin2022open,geva2000second,cresser2017coarse,albash2012quantum,mori2008dynamics}.  Equation (\ref{eq:non-canonical}) does not contain contributions from the real part of bath correlations because detailed balance condition cancel them. The condition $f(\omega)=0$ in Eq. (\ref{eq:def_f}) suggests $\mathcal{S} (-\omega) / \mathcal{S}(\omega) = -\exp (-\beta \omega) $ as a kind of detailed balance condition that leads the system dynamics to the Gibbs steady state. However, it is not satisfied in general. Non-Gibbs steady state for the system due to $\mathcal{S}(\omega)$ and non-secular contributions are even expected for stronger coupling strength between the oscillators, i.e., including counter-rating terms in the system Hamiltonian (details in Appendix \ref{appendix:non-canonical state}).

On the other hand, for Ohmic spectral densities in  Eq. (\ref{eq:Ohmic_a}) and (\ref{eq:Ohmic_b}), Eq. (\ref{eq:non-canonical}) shows that the steady state of the system differs from the Gibbs state when interacting oscillators A and B are unequally damped by their baths, i.e., $ \gamma_b - \gamma_a \neq 0$. The same conclusion was given in Ref. \cite{cresser2017coarse} for two interacting spins coupled to independent baths at thermal equilibrium. Meanwhile, the detuning dependence in Eq. (\ref{eq:non-canonical}) implies that the non-Gibbs state maximally differs from the Gibbs state in the resonant case ($\Delta=0$). In the dispersive regime $|\Delta|\gg g $, the system relaxes approximately in the Gibbs state. The oscillator damping and detuning dependences are a consequence of the non-secular terms in Eq. (\ref{eq:def_g_12}-\ref{eq:def_eta_21}).

In our model, the system steady state depends on system parameters, such as $\gamma_a$ and $\gamma_b$, in addition to the canonical temperature dependence in the Gibbs state (see Eq. (\ref{eq:canonical_state})). System steady states that depend on bath parameters have been reported using the mean force Gibbs state, $\hat{\rho}_S^{\rm (MFG)} \sim {\rm Tr}_B \left\{ \exp (-\beta \hat{\mathcal{H}}) \right\}$, which describes the system when the open quantum system is in thermal equilibrium. ${\rm Tr}_B$ is the trace over the bath degree of freedoms. The state $\hat{\rho}_S^{\rm (MFG)} \neq \hat{\rho}_G$ generally differs from the Gibbs state due to the weakly coupled bath \cite{cresser2021weak,trushechkin2022open}.

\subsection{Resonant deviation of oscillator occupations}

The non-secular terms couple the dynamics of normal mode occupations and normal modes coherence, as is observed in Appendix \ref{appendix:oscillator occupation}. It leads to steady values $\langle \hat{c}_1^\dagger \hat{c}_1 \rangle_{\rm ss} $ and $ \langle \hat{c}_2^\dagger \hat{c}_2 \rangle_{\rm ss}$ that deviate from their thermal values, $n^{\rm th}(\Omega_1)$ and $n^{\rm th}(\Omega_2)$, respectively, as a consequence of the system non-Gibbs steady state. Deviations are also observed in the steady occupation of oscillators A and B, $\langle \hat{a}^\dagger \hat{a} \rangle_{\rm ss}$ and $\langle \hat{b}^\dagger \hat{b} \rangle_{\rm ss}$, respectively. Under resonant condition and temperature regime $T \gtrsim g$, deviations of oscillator occupations, $\langle \hat{a}^\dagger \hat{a} \rangle_{\rm ss} - n^{\rm th}(\omega_b)\sim - {\rm Im}\{ \Gamma_{21} \}$ and $\langle \hat{b}^\dagger \hat{b} \rangle_{\rm ss} - n^{\rm th}(\omega_b) \sim {\rm Im}\{ \Gamma_{21} \}$, are due to imaginary part of bath correlations and non-secular terms in the dynamics, as for non-Gibbs relaxation in Eq. (\ref{eq:non-canonical}), through the dependence (derivation in Appendix \ref{appendix:compensation_relation})
\begin{eqnarray}\label{eq:Im_eta_21}
{\rm Im}\{ \Gamma_{21} \} = \frac{1}{2} \left( \gamma_b - \gamma_a \right) \left[  \mathcal{S}(-\Omega_1) - \mathcal{S}(-\Omega_2)  \right].
\end{eqnarray}
The term in Eq. (\ref{eq:Im_eta_21}) requires to evaluate $\mathcal{S}(\omega)$ in frequencies $\Omega_1 = \omega_b + g $ and $\Omega_2 = \omega_b -g$. It suggests that deviations of oscillator occupations are expected in the strong coupling regime due to the non-negligible Rabi splitting $\Omega_1 - \Omega_2 = 2g$. For weakly interacting oscillators, the condition $\Omega_1 \approx \Omega_2$ vanishes possible oscillator deviations from Eq. (\ref{eq:Im_eta_21}) because ${\rm Im}\{ \Gamma_{21} \} \rightarrow 0$.

Under resonant condition and $T \gtrsim g$, complete expressions for steady oscillator occupations (details in Appendix \ref{appendix:oscillator occupation}),
\begin{eqnarray}
\langle \hat{a}^\dagger \hat{a} \rangle_{\rm ss} - n^{\rm th}(\omega_b)  & \approx&  - \frac{2g}{\gamma_a}{\rm Im} \langle \hat{a} \hat{b}^\dagger \rangle_{\rm ss},\label{eq:approx_resonant_occupation_a}\\
\langle \hat{b}^\dagger \hat{b} \rangle_{\rm ss} - n^{\rm th}(\omega_b) &\approx &  \frac{2g}{\gamma_b}{\rm Im} \langle \hat{a} \hat{b}^\dagger \rangle_{\rm ss},\label{eq:approx_resonant_occupation_b}
\end{eqnarray}
can be written in terms of the non-zero steady coherence between oscillators A and B, $\langle \hat{a} \hat{b}^\dagger \rangle_{\rm ss}$, associated to the non-Gibbs state. The same occupation-coherence dependence in Eq. (\ref{eq:approx_resonant_occupation_a}) and (\ref{eq:approx_resonant_occupation_b}) was derived for a local-Lindblad model in Ref. \cite{ahn2023modification}, and was applied to explain cavity-modified vibrational occupations under the resonant condition. Therefore, Eq. (\ref{eq:approx_resonant_occupation_a}) and (\ref{eq:approx_resonant_occupation_b}) suggest connections between steady Redfield and local-Lindblad approaches.

\begin{figure}[t!]
\centering
\includegraphics[width=0.48\textwidth]{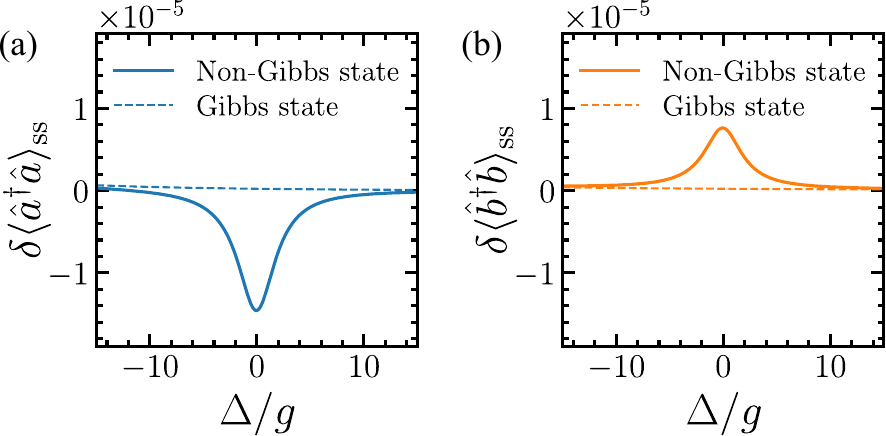}
\caption{Steady occupations of strongly interacting oscillators A and B as a function of the detuning $\Delta$. Panels (a) and (b) correspond to values of $\delta \langle \hat{a}^\dagger \hat{a} \rangle_{\rm ss} = \langle \hat{a}^\dagger \hat{a} \rangle_{\rm ss} - n^{\rm th}(\omega_a)$ and $\delta \langle \hat{b}^\dagger \hat{b} \rangle_{\rm ss} = \langle \hat{b}^\dagger \hat{b} \rangle_{\rm ss} - n^{\rm th}(\omega_b)$, respectively. System parameters are $\{ \gamma_a,\gamma_b,T,\omega_b,\omega_c  \} = \{ 10^{-2},2\times 10^{-2}, 10,10^2,10^3  \}g$.}
\label{fig:non_canonical_occupations}
\end{figure}

We verify the expected deviations of oscillator occupations by computing $\delta \langle \hat{a}^\dagger \hat{a} \rangle_{\rm ss} \equiv \langle \hat{a}^\dagger \hat{a} \rangle_{\rm ss} - n^{\rm th}(\omega_a)$ and $\delta \langle \hat{b}^\dagger \hat{b} \rangle_{\rm ss} \equiv \langle \hat{b}^\dagger \hat{b} \rangle_{\rm ss} - n^{\rm th}(\omega_b)$ as a function of $\Delta$, shown in Fig. \ref{fig:non_canonical_occupations} (a) and (b), respectively. The results show that in the non-Gibbs state the oscillator occupations resonantly deviates from thermal around to $10\%$ of the thermal values, increasing the occupation of oscillator A in contrast to oscillator B. Under the resonant condition, the fact that effect on A is double of the effect on B is explained by the balance condition $\gamma_a \langle \hat{a}^\dagger \hat{a} \rangle_{\rm ss} +  \gamma_b \delta \langle \hat{b}^\dagger \hat{b} \rangle _{\rm ss} \approx 0 $ that merges from mixing Eq. (\ref{eq:approx_resonant_occupation_a}) and (\ref{eq:approx_resonant_occupation_b}). The balance condition means that one oscillator increases the occupation while the other decreases the occupation in a proportion of $\gamma_a/\gamma_b$, which is a factor of $2$ in our case, and merges as a consequence of the conservation of the number of excitations in the open quantum system, as we will see later. When $|\Delta |>g$, Fig. \ref{fig:non_canonical_occupations} shows oscillator occupations that are similar to thermal values due to the system steady state being approximately the Gibbs state (see Eq. (\ref{eq:non-canonical})). All in all, the observed resonant feature of oscillator occupations is a signature of the non-Gibbs state of the system and merges as a consequence of $ \langle \hat{a} \hat{b}^\dagger \rangle_{\rm ss}$.

\subsection{Unbalanced oscillator states}
\begin{figure}[t!]
\centering
\includegraphics[width=0.42\textwidth]{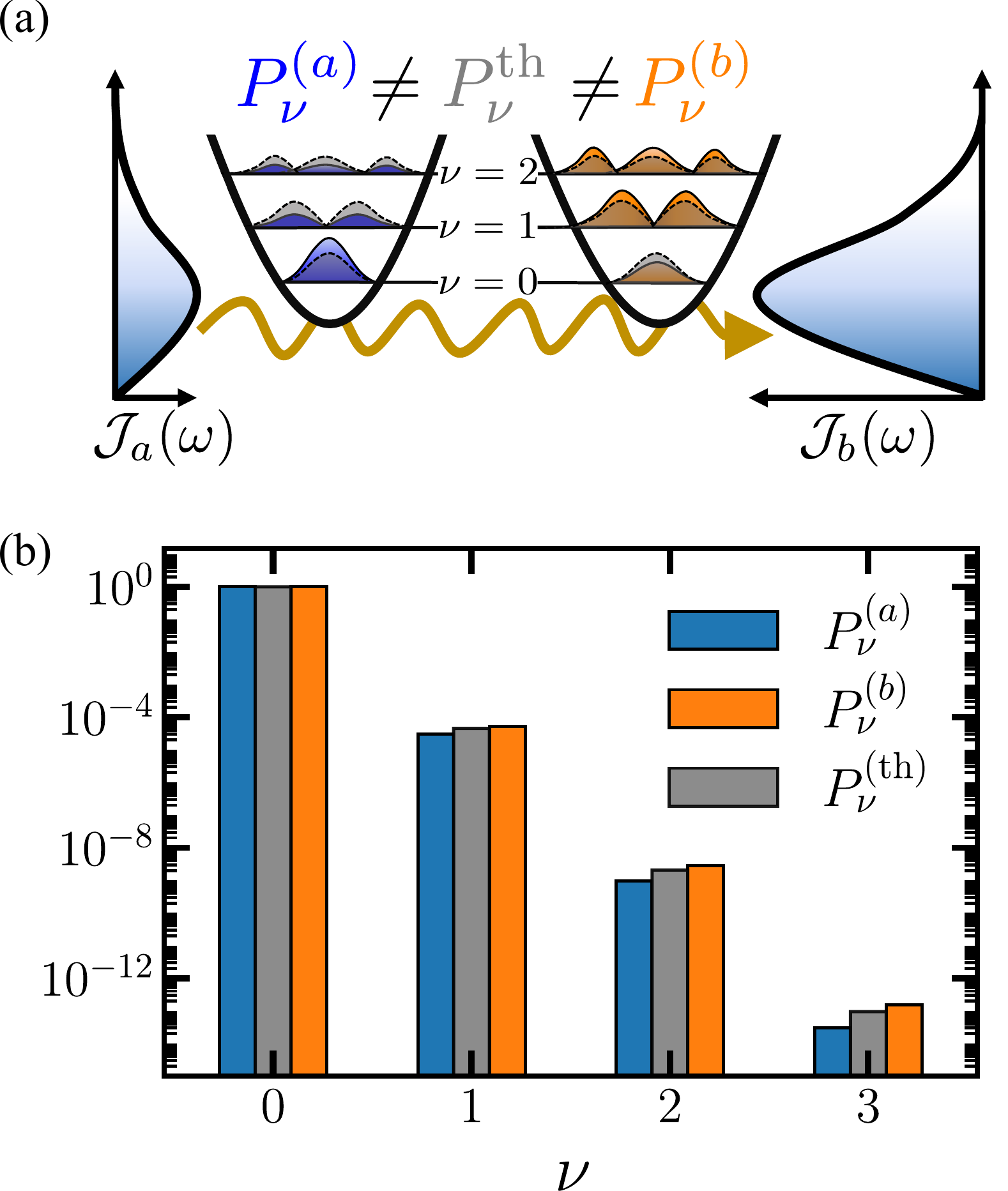}
\caption{(a) Scheme of unbalanced probability distributions of resonant strongly interacting oscillators under steady flux of excitations (yellow arrow) when damping rates are $\gamma_a < \gamma_b$. (b) Excitation probability densities $P_\nu^{(a)}$ and $P_\nu^{(b)}$ over Fock states $\nu$ for resonant oscillators in the non-Gibbs states, compared to the thermal probability distribution $P_\nu^{\rm th}$.} 
\label{fig:probability}
\end{figure}
We complement the analysis of oscillator occupations by analyzing the state of oscillators A and B, $\hat{\rho}_a $ and $ \hat{\rho}_b$, respectively. They are constructed by tracing the system non-Gibbs state with respect to the remaining oscillator. As illustrated in Fig. \ref{fig:probability}(a), the excitation of oscillator A is distributed in the Fock states $\ket{\nu} = \{ \ket{0},\ket{1},\ket{2},... \}$ according to the probability distribution $P_\nu^{(a)} = \bra{\nu } \hat{\rho}_a \ket{\nu}$. $P_\nu^{(b)}$ gives an equivalent description for oscillator B.  The distributions are compared with the probability distribution of a thermal equilibrium oscillator with frequency $\omega_b$ at temperature $T$, $P_\nu^{(\rm th)}= \left[ 1-\exp(-\beta\omega_b) \right] \exp(-\beta\omega_b \nu)$.

Figure \ref{fig:probability} (b) shows the probability distributions of the strongly interacting oscillators under resonant condition. The results show that our obtained non-Gibbs state gives positive probabilities, as a consequence of the partial secular approximation employed and the validity of weak system-bath coupling in our parameters \cite{hartmann2020accuracy}. With respect to $P_\nu^{(\rm th)}$, oscillator A has a lower probability of being in the excited levels ($\nu > 0$) in opposition to oscillator B, as illustrated in Figure \ref{fig:probability}(a). This means that resonant oscillators have distributions, $P_\nu^{(a)}\neq P_\nu ^{(\rm th)} \neq P_\nu^{(b)}$, that are not equilibrated ({\it unbalanced}) due to the condition $\gamma_a \neq \gamma_b$. These probability distributions consistently describe the resonant feature of the oscillator occupations observed in Fig. \ref{fig:non_canonical_occupations}, where the highly damped oscillator is more excited in contrast to the lightly damped oscillator. Unbalanced states are neither expected nor observed in the Gibbs state.

\subsection{Resonant excitation flux}
Based on the dynamics of normal mode occupations (details in Appendix \ref{appendix:oscillator occupation}), the dynamics of the mean number of excitations in the bath dynamics, in Eq. (\ref{eq:bath_particles_a}) and (\ref{eq:bath_particles_b}), leads to the relation
\begin{equation}\label{eq:bath-system_relation}
\begin{split}
\frac{\rm d}{\rm dt} \langle \hat{\mathcal{N}}_a \rangle + \frac{\rm d}{\rm dt} \langle \hat{\mathcal{N}}_b \rangle + \frac{\rm d}{\rm dt} \langle \hat{\mathcal{N}}_S \rangle =0,
\end{split}
\end{equation}
where $\hat{\mathcal{N}}_S = \hat{c}_1^\dagger \hat{c}_1 + \hat{c}_2^\dagger \hat{c}_2$ is the number operator of excitations in the system. Equation (\ref{eq:bath-system_relation}) shows that the constructed partial secular Redfield approach conserves the mean number of excitations of the open quantum system at any time and over all whole range of parameters. In the steady state, the system condition $\frac{\rm d}{\rm dt} \langle \hat{\mathcal{N}}_S \rangle_{\rm ss} =0$ in Eq. (\ref{eq:bath-system_relation}) implies that any possible excitation that one of the bath releases to the system should be transferred to the other bath, generating a steady flux of excitations, $\frac{\rm d}{\rm dt} \langle \hat{\mathcal{N}}_a \rangle_{\rm ss} = -  \frac{\rm d}{\rm dt} \langle \hat{\mathcal{N}}_b \rangle_{\rm ss}$, which conserves the total number of particles. 

When the strongly interacting oscillators are resonant and $T \gtrsim g$, the approximations scheme employed to derive Eq. (\ref{eq:approx_resonant_occupation_a}) and (\ref{eq:approx_resonant_occupation_b}) reduces the steady bath dynamics in Eq. (\ref{eq:bath_particles_a}) and (\ref{eq:bath_particles_b}) to
\begin{eqnarray}\label{eq:resonant_current}
    \frac{\rm d}{\rm dt} \langle \hat{\mathcal{N}}_a \rangle_{\rm ss} = -\frac{\rm d}{\rm dt} \langle \hat{\mathcal{N}}_b \rangle_{\rm ss} \approx  -2g {\rm Im} \langle \hat{a} \hat{b}^\dagger \rangle_{\rm ss}.
\end{eqnarray}
In steady state, Eqs. (\ref{eq:resonant_current}), (\ref{eq:approx_resonant_occupation_a}) and (\ref{eq:approx_resonant_occupation_b}) with the parameters used in Fig. \ref{fig:non_canonical_occupations} ($\gamma_b>\gamma_a$) imply that the bath connected to oscillator A is constantly decreasing in excitation number, while bath increases. Therefore, the system experiences a excitation flux from small to high oscillator damping, as illustrated in Fig. \ref{fig:probability}(a). The steady excitation flux is maximum under resonant condition, as the stationary coherence ${\rm Im} \langle \hat{a} \hat{b}^\dagger \rangle_{\rm ss}$ is a maximum on resonance. The stationary oscillator coherence is associated with the long-time formation of a non-Gibbs state. The Gibbs state gives zero steady flux of excitations in all the regime of parameters. The steady flux is a resonant non-Gibbs state feature that vanishes in the dispersive regime $|\Delta| \gg g$ because the system relaxes to the Gibbs state in this limit.

\section{Conclusion and Outlook}\label{sec:conclusion}

In this work, we study the steady state of two strongly interacting oscillators, each of them weakly coupled to independent baths at the same temperature. The derived system dynamics was constructed using the partial secular Redfield approach, which gives positive probabilities and preserves non-negligible non-secular terms that couple the system normal modes when oscillators are unequally damped by their baths. Theoretical analysis shows that the imaginary part of bath correlations and non-secular terms are responsible for the non-Gibbs state relaxation, as was suggested in Ref. \cite{geva2000second,cresser2017coarse,albash2012quantum,mori2008dynamics}.

The results show that oscillators relax in the non-Gibbs state when they are nearly resonant. Specifically, in the resonant condition, their steady excitation distributions and occupations deviate from thermal values. The deviation is positive for one of the oscillators and negative for the other following a balance condition such that the total number of excitations in the open quantum system is conserved. The unequal damping rates establish a preferential direction for the physical feature of the non-Gibbs state, controlling the deviation of oscillator occupations and generating a flux of excitations from the lightly to the highly damped oscillator. The system dynamics relaxes to the usual Gibbs state when oscillators are large detuned or weakly interacting oscillators.

Approximate expressions for system and bath observables show that the non-zero steady coherence between the oscillators in the non-Gibbs state is responsible for the resonant modification of steady oscillator occupations and the steady flux of excitations. This argument coincides with the local-Lindblad approach, where the steady light-matter coherence is responsible for the observed modified vibrational occupations \cite{ahn2023modification}. It highlights non-Gibbs steady states as a feasible mechanism to describe reported apparent modifications of thermal occupations of strongly interacting quantum subsystems in thermal equilibrium \cite{yu2023ostensible,ahn2023modification}.

\begin{acknowledgments}
F.R. is supported by ANID Doctoral Scholarship 21221970, and A.E.R.L. and F.H. are supported by ANID through grants FONDECYT Iniciaci\'on No. 11250638, FONDECYT Regular No. 1221420 and the Millennium Science Initiative Program ICN17\_012. This work was supported by the international collaborative grant ANID/ECOS-Sud C20E01 (``Quantum dynamics in cavity-coupled molecules''). J.S.~is supported by the ERC Consolidator project MATHLOCCA (Grant nr.~101170485) and by the French National Research Agency under the France 2030 program ANR-23-PETQ-0002 (PEPR project QUTISYM) and project ANR-21-ESRE-0032 (PEPR project aQCess). 
\end{acknowledgments}

\appendix

\section{Derivation of the Redfield master equation}\label{appendix:Redfield_QME}

Here we derive the Redfield master equation 
\begin{eqnarray}\label{eq:Redfield_equation}
    \frac{\rm d}{\rm dt} \hat{\rho}_S &=& -i \left[ \hat{\mathcal{H}}_S,\hat{\rho}_S \right] \\
    &-&\int_0^\infty {\rm d} \tau {\rm Tr}_B \Big\{ \left[\hat{\mathcal{H}}_{SB},\left[\hat{\mathcal{H}}_{SB}(-\tau),\hat{\rho}_S\otimes \hat{\rho}_B\right] \right] \Big\}, \nonumber
\end{eqnarray}
for our system described in Sec. \ref{sec:Hamiltonians}. In the usual microscopic derivation, the system-bath interactions in Eq. (\ref{eq:interaction_Hamiltonian}), $\hat{\mathcal{H}}_{SB} =  \sum_\alpha \hat{q}_\alpha  \otimes \hat{Q}_\alpha$, is written as a sum of individual system-bath interactions of oscillator A ($\alpha = a$) and B ($\alpha = b$). The bi-linear structure of the interaction is composed of linear system operators, $\hat{q}_a = (\hat{a} + \hat{a}^\dagger)$ and $\hat{q}_b = (\hat{b} + \hat{b}^\dagger)$, and linear bath operators, $\hat{\mathcal{Q}}_a = \sum_i \lambda_i^{(a)} (\hat{a}_i + \hat{a}_i^\dagger)$ and $\hat{\mathcal{Q}}_b = \sum_j \lambda_j^{(b)} (\hat{b}_j + \hat{b}_j^\dagger)$. Using the previous decomposition, we write the interaction Hamiltonian in the interaction picture, $\hat{\mathcal{H}}_{SB}(t) = \hat{\mathcal{U}}_t \hat{\mathcal{H}}_{SB} \hat{\mathcal{U}}_t^\dagger= \sum_\alpha \hat{q}_\alpha(t) \otimes \hat{\mathcal{Q}}_\alpha(t) $, in terms of time-dependent system and bath operators, $\hat{q}_\alpha(t) = \hat{\mathcal{U}}_t \hat{q}_\alpha \hat{\mathcal{U}}_t^\dagger $ and $\hat{\mathcal{Q}}_\alpha(t) = \hat{\mathcal{U}}_t \hat{\mathcal{Q}}_\alpha \hat{\mathcal{U}}_t^\dagger$, respectively. The unitary transformation $\hat{\mathcal{U}}_t= \exp(i [\hat{\mathcal{H}}_S + \hat{\mathcal{H}}_B]t)$ moves from Schrodinger to the interaction picture. In terms of the previous system and bath operators, the Redfield dynamics in Eq. (\ref{eq:Redfield_equation}) is reduced to the structure \cite{davidovic2020completely,scali2021local}
\begin{equation}\label{eq:def_Redfield_QME}
\frac{\rm d}{\rm dt} \hat{\rho}_S = -i \Big[ \overline{\mathcal{H}}_S,\hat{\rho}_S  \Big] +  \mathcal{D}[\hat{\rho}_S].
\end{equation}
We check that the system dynamics in Eq. (\ref{eq:def_Redfield_QME}) conserves the system probability, because $\frac{\rm d}{\rm dt}{\rm Tr}_S\{  \hat{\rho}_S \}=0$. In Eq. (\ref{eq:def_Redfield_QME}), the system is subject to coherent evolution given by 
\begin{equation}\label{eq:def_effective_Hamiltonian}
\overline{\mathcal{H}}_S = \hat{\mathcal{H}}_S +\frac{i}{2} \sum_{\alpha = \{a,b\}} \left( \hat{r}_\alpha^\dagger \hat{q}_\alpha - \hat{q}_\alpha \hat{r}_\alpha \right),
\end{equation}
which is an Hermitian Hamiltonian that includes modifications induced by the baths. Equation (\ref{eq:def_effective_Hamiltonian}) is written in terms of the system operators
\begin{equation}\label{eq:def_system_integrals}
\hat{r}_\alpha = \int_0^\infty  {\rm d} \tau \hat{q}_\alpha (-\tau) \langle \hat{\mathcal{Q}}_\alpha \hat{\mathcal{Q}}_\alpha(-\tau) \rangle.
\end{equation}
that encodes the bath dependencies through the bath correlations $\langle \hat{\mathcal{Q}}_\alpha \hat{\mathcal{Q}}_\alpha(-\tau) \rangle = {\rm Tr}_B \{ \hat{\mathcal{Q}}_\alpha \hat{\mathcal{Q}}_\alpha(-\tau) \hat{\rho}_B \}$. Meanwhile, the dissipation of the system in Eq. (\ref{eq:def_Redfield_QME}) is described by
\begin{equation}\label{eq:def_dissipation}
\mathcal{D} [\hat{\rho}_S] =  \sum_{\alpha = \{a,b\}} \left(  \hat{r}_\alpha \hat{\rho}_S \hat{q}_\alpha -\frac{1}{2}\left\{  \hat{q}_\alpha \hat{r}_\alpha ,\hat{\rho}_S \right\} + {\rm h.c.} \right),
\end{equation}
where $\rm h.c.$ denotes the Hermitian conjugated. 

\subsection{Dynamics in the normal mode basis}

To explicitly derive $\hat{r}_a$ and $\hat{r}_b$ in Eq. (\ref{eq:def_system_integrals}), we write the system Hamiltonian in diagonal form by considering the Bogoliubov transformation \cite{xiao2009theory}
\begin{eqnarray}
\hat{a} = \sum_{i=\{ 1,2 \}}  \left( U_{a,i} \hat{c}_i + V_{a,i} \hat{c}_i^\dagger \right),\label{eq:Bogoliubov_CR_a} \\
\hat{b} = \sum_{i=\{ 1,2 \}}  \left( U_{b,i} \hat{c}_i + V_{b,i} \hat{c}_i^\dagger \right), \label{eq:Bogoliubov_CR_b}
\end{eqnarray}
where coefficients $ U_{\alpha,i}$ and $V_{\alpha,i}$ are conveniently chosen such that $\hat{\mathcal{H}}_S$ in Eq. (\ref{eq:system_Hamiltonian}) is reduced to the two normal modes, such as in Eq. (\ref{eq:diagonal_Hamiltonian}). $\hat{c}_i$ corresponds to the normal mode anihilation operators with $i=\{ 1,2 \}$ the normal mode indexes.

Under the transformations in Eq. (\ref{eq:Bogoliubov_CR_a}) and (\ref{eq:Bogoliubov_CR_b}), system operators $\hat{q}_a =(\hat{a} + \hat{a}^\dagger)$ and $\hat{q}_b= (\hat{b} + \hat{b}^\dagger)$ in the normal mode basis reduces to
\begin{equation}\label{eq:system_operators}
\hat{q}_\alpha= \sum_{i=\{ 1,2 \}} \left(   \chi_{\alpha,i} \hat{c}_i + \chi_{\alpha,i}^* \hat{c}_i^\dagger \right),
\end{equation}
whit $\chi_{\alpha ,i} = U_{\alpha ,i} + V_{\alpha ,i}^*$ coefficients. Equation (\ref{eq:system_operators}) in the interaction picture is directly reduced to
\begin{equation}\label{eq:system_operators_tau}
\hat{q}_\alpha(t) = \sum_{i=\{ 1,2 \}} \left(   \chi_{\alpha,i} e^{-i\Omega_i t}\hat{c}_i + \chi_{\alpha,i}^* e^{i\Omega_i t} \hat{c}_i^\dagger \right).
\end{equation}
Meanwhile, due to the thermal state of the baths, bath correlations are reduced to
\begin{eqnarray}\label{eq:second_bath}
\langle \hat{\mathcal{Q}}_\alpha \hat{\mathcal{Q}}_\alpha(-\tau) \rangle &=&\gamma_\alpha \int_{0}^{\infty} {\rm d} \omega \mathcal{J} (\omega)  \cos(\omega \tau)  \coth\left( \frac{ \beta \omega}{2} \right) \nonumber \\
&-&i\gamma_\alpha\int_{0}^{\infty}  {\rm d} \omega  \mathcal{J} (\omega)  \sin(\omega \tau),
\end{eqnarray}
in terms of the bath spectral densities $\mathcal{J}_a (\omega)=\gamma_a \mathcal{J}(\omega)$ and $\mathcal{J}_b (\omega)= \gamma_b \mathcal{J}(\omega)$ according to Eq. (\ref{eq:Ohmic_a}) and (\ref{eq:Ohmic_b}), respectively.

Using the system and bath terms in Eq. (\ref{eq:system_operators_tau}) and (\ref{eq:second_bath}), respectively, system operators in Eq. (\ref{eq:def_system_integrals}) have the structure
\begin{equation}\label{eq:system_integral}
\hat{r}_\alpha = \gamma_\alpha \sum_{i=\{ 1,2 \}} \left(   \chi_{\alpha,i} \mathcal{C} (\omega_i)\hat{c}_i + \chi_{\alpha,i}^* \mathcal{C} (-\omega_i)\hat{c}_i^\dagger \right),
\end{equation}
in terms of the bath function
\begin{equation}\label{eq:def_bath_correlations}
    \mathcal{C}(\omega) =\frac{1}{\gamma_\alpha} \int_0^\infty d \tau \langle \hat{\mathcal{Q}}_\alpha \hat{\mathcal{Q}}_\alpha(-\tau) \rangle \exp(i\omega \tau).
\end{equation}
Using the formula
\begin{equation}
\int_0^{\infty} d\tau e^{\pm i\omega \tau} = \pi \delta(\omega) \pm i \mathcal{P} \frac{1}{\omega},
\end{equation}
the function in Eq. (\ref{eq:def_bath_correlations}) is reduced to Eq. (\ref{eq:bath_correlations}), whit $\mathcal{P}$ the Cauchy principal value.

Under the normal mode description of the system (see Eq. (\ref{eq:system_operators}) and (\ref{eq:system_integral})), the Hamiltonian in Eq. (\ref{eq:def_effective_Hamiltonian}) can be reduced to
\begin{equation}\label{eq:effective_Hamiltonian}
\overline{\mathcal{H}}_S  = \sum_i \overline{\Omega}_i \hat{c}_i^\dagger \hat{c}_i +  \sum_{i,j\neq i} h_{i,j}^{(\rm s)} \hat{c}_i^\dagger \hat{c}_j +  \sum_{i,j} \left( h_{i,j}^{(\rm f)} \hat{c}_i^\dagger \hat{c}_j^\dagger + {\rm h.c.} \right). 
\end{equation}
The first term in Eq. (\ref{eq:effective_Hamiltonian}) describes bath induced modifications of the normal-mode frequencies (Lamb-shifts), $\overline{\Omega}_i = \Omega_i + \sum_\alpha \gamma_\alpha |\chi_{\alpha , i}|^2 \left[  \mathcal{S} (\omega) + \mathcal{S} (-\omega) \right]$, as linear combinations of $\mathcal{S}(\omega)$ in Eq. (\ref{eq:def_lamb_shifts}). The second term in Eq. (\ref{eq:effective_Hamiltonian}) describes the energy exchange between normal modes, with a coupling strengths
\begin{eqnarray}\label{eq:slow_energies}
h_{i,j}^{(\rm s)}  =  \frac{i}{2} \sum_{\alpha = \{a,b \}} \gamma_\alpha\chi_{\alpha , i}  ^* \chi_{\alpha , j} \Big[ \mathcal{C} (\Omega_i)^* &-&\mathcal{C} (\Omega_j) \nonumber \\
+  \mathcal{C} (-\Omega_j)^* &-& \mathcal{C} (-\Omega_i) \Big],
\end{eqnarray}
in terms of $\mathcal{C}(\omega)$ in Eq. (\ref{eq:bath_correlations}). The third term in Eq. (\ref{eq:effective_Hamiltonian}) describes normal mode squeezing when $j=i$, and double creation and annihilation in the normal modes when $j\neq i$, with the corresponding strength
\begin{equation}\label{eq:fast_energies}
h_{i,j}^{(\rm f)}  =  \frac{i}{2} \sum_{\alpha = \{a,b \}} \gamma_\alpha\chi_{\alpha , i}^* \chi_{\alpha , j}^*  \left[ \mathcal{C}(\Omega_i)^* -  \mathcal{C} (-\Omega_j)  \right].
\end{equation}

On the other hand, system operators in Eq. (\ref{eq:system_operators}) and (\ref{eq:system_integral}) reduce the dissipative term in Eq. (\ref{eq:def_dissipation}) to
\begin{eqnarray}\label{eq:dissipation}
    &&\mathcal{D}[\hat{\rho}_S] = \sum_i \Big( \Gamma_i (\Omega_i) \mathcal{L}_{\hat{c}_i} [\hat{\rho}_S]  +\Gamma_i (-\Omega_i) \mathcal{L}_{\hat{c}_i^\dagger} [\hat{\rho}_S]  \Big)\nonumber \\
    &&+\sum_{i,j \neq i} \Big( \Gamma_{i,j}^{(\rm s)}(\Omega_i,\Omega_j) \mathcal{L}_{\hat{c}_i^\dagger, \hat{c}_j} [\hat{\rho}_S] 
    + \Gamma_{i,j}^{(\rm s)}(-\Omega_j, - \Omega_i)^* \mathcal{L}_{\hat{c}_i, \hat{c}_j^\dagger} [\hat{\rho}_S] \Big)\nonumber \\
    &&+\sum_{i,j} \Big( \Gamma_{i,j}^{(\rm f)} \mathcal{L}_{\hat{c}_i^\dagger, \hat{c}_j^\dagger} [\hat{\rho}_S] + \Gamma_{i,j}^{(\rm f)*} \mathcal{L}_{\hat{c}_j, \hat{c}_i} [\hat{\rho}_S] \Big).
\end{eqnarray}
The first term in Eq. (\ref{eq:dissipation}) describes excitation and de-excitation of the normal-mode  in the Lindblad form (see Eq. (\ref{eq:Lindblad})) at rates $\Gamma_i (\Omega_i)$, and $\Gamma_i (-\Omega_i)$, respectively. Rates $\Gamma_i (\omega) =  \sum_\alpha \gamma_\alpha|\chi_{\alpha,i}|^2 \Gamma(\omega)$ are linear combinations of $\Gamma(\omega)$ in Eq. (\ref{eq:def_rates}). The second and third terms in Eq. (\ref{eq:dissipation}) describe mix dissipation between normal modes when $j \neq i$, and dissipative effects in each normal mode when $i=j$. These terms are in the general structure in Eq. (\ref{eq:cross_Lindblad}), and have the corresponding rates
\begin{eqnarray}
\Gamma_{i,j}^{(\rm s)} (\Omega, \Omega') =   \sum_{\alpha = \{a,b \}} \gamma_\alpha \chi_{\alpha , i}  ^* \chi_{\alpha , j} \Big[ \mathcal{C} (\Omega)^* +  \mathcal{C} (\Omega')  \Big],\label{eq:slow_rates}\\
\Gamma_{i,j}^{(\rm f)}  =  \sum_{\alpha = \{a,b \}} \gamma_\alpha \chi_{\alpha , i}^*   \chi_{\alpha , j}^* \Big[ \mathcal{C} (\Omega_i)^* + \mathcal{C} (-\Omega_j)  \Big].\label{eq:fast_rates}
\end{eqnarray}

Based on Eq. (\ref{eq:effective_Hamiltonian}) and (\ref{eq:dissipation}), the Redfield master equation in Eq. (\ref{eq:def_Redfield_QME}) is finally reduced to
\begin{widetext}
\begin{eqnarray}\label{eq:full_Redfield}
\frac{\rm d}{\rm dt} \hat{\rho}_S &=& \sum_i \Big\{-i \overline{\Omega}_i \left[  \hat{c}_i^\dagger \hat{c}_i ,  \hat{\rho}_S  \right] +   \Gamma_i (\Omega_i) \mathcal{L}_{\hat{c}_i} [\hat{\rho}_S]  +\Gamma_i (-\Omega_i) \mathcal{L}_{\hat{c}_i^\dagger} [\hat{\rho}_S] \Big\}   \nonumber \\
&+& \sum_{i,j\neq i} \Big\{-i h_{i,j}^{(\rm s)} \left[  \hat{c}_i^\dagger \hat{c}_j , \hat{\rho}_S \right] +    \Gamma_{i,j}^{(\rm s)}(\Omega_i,\Omega_j) \mathcal{L}_{\hat{c}_i^\dagger, \hat{c}_j} [\hat{\rho}_S] 
+ \Gamma_{i,j}^{(\rm s)}(-\Omega_j, - \Omega_i)^* \mathcal{L}_{\hat{c}_i, \hat{c}_j^\dagger} [\hat{\rho}_S]  \Big\} \nonumber \\
&+& \sum_{i,j}\Big\{ -i  h_{i,j}^{(\rm f)} \left[ \hat{c}_i^\dagger \hat{c}_j^\dagger, \hat{\rho}_S \right] + \Gamma_{i,j}^{(\rm f)} \mathcal{L}_{\hat{c}_i^\dagger, \hat{c}_j^\dagger} [\hat{\rho}_S] -i h_{i,j}^{(\rm f)*} \left[ \hat{c}_i\hat{c}_j, \hat{\rho}_S \right]   + \Gamma_{i,j}^{(\rm f)*} \mathcal{L}_{\hat{c}_j, \hat{c}_i} [\hat{\rho}_S] \Big\}.
\end{eqnarray}
\end{widetext}
\subsection{Partial secular approximation}\label{appendix:partial_secular}

When we write the Redfield master equation in Eq. (\ref{eq:full_Redfield}) in the interaction picture, the evolution of $\hat{\rho}_S (t) = \hat{\mathcal{U}}_t \hat{\rho }_S\hat{\mathcal{U}}_t^\dagger$ contains time-independent (secular) and time-dependent (non-secular) terms. The non-secular terms oscillate with frequencies $\omega_0 = \{ 2\Omega_1 , 2\Omega_2, \Omega_1 - \Omega_2 , \Omega_1 + \Omega_2 \}$. In order to prevent negative system probabilities that emerge from fast oscillating non-secular terms, here we decrease the impact of part of the non-secular terms by deriving a coarse-grained system dynamics \cite{hartmann2020accuracy,schaller2008preservation}. This implies integrating the evolution of $\hat{\rho}_S (t)$ over a time window $T_0$, reducing secular and non-secular terms to \cite{davidovic2020completely}
\begin{eqnarray}
\int_{t-T_0/2}^{t+T_0/2} d\tau \hat{\rho}_S({\tau}) &\approx & \hat{\rho}_S(t), \label{eq:CG_approximation_sec}\\
\int_{t-T_0/2}^{t+T_0/2} d\tau e^{i\omega_0 \tau} \hat{\rho}_S({\tau}) &\approx & e^{i\omega_0 t}\hat{\rho}_S(t) {\rm sinc}\Bigg(\frac{\omega_0 T_0 }{2} \Bigg), \nonumber\\
\label{eq:CG_approximation_non-sec}
\end{eqnarray}
respectively. Previous approximated expressions assume that $\hat{\rho}_S(t)$ behaves as constant over the time window, which is feasible for $T_0\ll1/ \{ \gamma_a,\gamma_b \}$ much smaller than the system relaxation time \cite{davidovic2020completely}. In the Schrodinger picture, the non-secular terms of the coarse-grained Redfield master equation are weighted by the factors ${\rm sinc} (\omega_0 T_0/2)$. Therefore, the impact of each non-secular term depends on its oscillating frequencies, being conserved (${\rm sinc} (\omega_0 T_0/2) \approx 1$) the slow oscillating ones $\omega_0 <1/T_0$, and being vanished (${\rm sinc} (\omega_0 T_0/2) \approx 0$) the fast oscillating ones $\omega_0 \gg 1/T_0$.

In our regime of parameters, we assume normal mode frequencies $ \Omega_i \gg  \gamma_\alpha$ much larger than the oscillator damping rates, and the normal mode frequency difference $|\Omega_1 - \Omega_2| < 1/T_0$ smaller than $1/T_0$. For integration time $T_0\ll 1/ \{ \gamma_a,\gamma_b \}$, previous assumptions establish the condition
\begin{equation}\label{eq:partial-secular_condition}
 \gamma_\alpha T_0 < |\Omega_1 - \Omega_2| T_0 < 1 \ll ( \Omega_i + \Omega_j )T_0,
\end{equation}
for $\alpha= \{a,b \}$ and $\{ i,j\} = \{ 1,2 \}$. Equation (\ref{eq:partial-secular_condition}) justifies vanishing fast oscillating non-secular terms from the coarse-grained dynamics, while slowly oscillating non-secular terms associated with the frequency $\Omega_1 - \Omega_2$ are preserved. Therefore, we obtain a partial secular Redfield master equation, where the fast oscillating terms with sub-index $\rm (f)$ in Eq. (\ref{eq:full_Redfield}) are ignored.

For strongly interacting oscillators that are nearly resonant ($|\Delta|<g$), normal mode frequencies in  Eq. (\ref{eq:omega_1}) and (\ref{eq:omega_2}) reduce to $\Omega_1 \approx \omega_b + g $ and $\Omega_2 \approx \omega_b - g$, respectively. These frequencies fulfill the condition in Eq. (\ref{eq:partial-secular_condition}) in our parametric analysis regime $\gamma_\alpha \ll g \ll \omega_\alpha$, reducing Eq. (\ref{eq:full_Redfield}) to Eq. (\ref{eq:CG_Redfield}). In the strong coupling regime, the corresponding factors $\{ \chi_{a,1},\chi_{a,2},\chi_{b,1},\chi_{b,2} \} = \{ \cos(A),-\sin(A),\sin(A),\cos(A) \}$ leads to the master equation parameters $\overline{g} = h_{2,1}^{\rm (s)}$, $\Gamma_{12} = \Gamma_{2,1}^{\rm (s)}(\Omega_1, \Omega_2)$  and $\Gamma_{21} = \Gamma_{2,1}^{\rm (s)}(-\Omega_1, - \Omega_2)^* $ defined in Eq. (\ref{eq:slow_energies}) and (\ref{eq:slow_rates}).

\subsection{Dynamics of mean number of excitations in the baths}\label{appendix:conservation of particles}

Here we derive expressions for the bath dynamics using the Redfield approach developed in Appendix \ref{appendix:Redfield_QME} for strongly interacting oscillators. The dynamics of the mean number of excitations in the baths connected to the oscillator A ($\alpha=a$) and B ($\alpha=b$) is given by \cite{recabal2026}
\begin{eqnarray}\label{eq:def_bath_particles}
    &&\frac{\rm d}{\rm dt} \langle \hat{\mathcal{N}}_\alpha \rangle = \nonumber \\
    &&-\int_0^\infty {\rm d} \tau {\rm Tr}  \Big\{  \hat{\mathcal{N}}_\alpha \left[ \hat{\mathcal{H}}_{SB}, \left[ \hat{\mathcal{H}}_{ SB}(-\tau) , \hat{\rho}_S \otimes \hat{\rho}_B \right] \right] \Big\}, \hspace{0.8cm}
\end{eqnarray}
where $\hat{\mathcal{N}}_\alpha$ is the number operator of excitations. Using the structure $\hat{\mathcal{H}}_{SB} =  \sum_\alpha \hat{q}_\alpha  \otimes \hat{\mathcal{Q}}_\alpha$ for the system-bath interaction Hamiltonian, Eq. (\ref{eq:def_bath_particles}) reads
\begin{eqnarray}\label{eq:flux_2}
\frac{\rm d}{\rm dt} \langle \hat{\mathcal{N}}_\alpha \rangle  = -\int_0^\infty {\rm d}\tau \Big\{  \bigl<  \hat{q}_\alpha \hat{q}_\alpha (-\tau) \bigr>  \bigl< [ \hat{\mathcal{Q}}_\alpha , \hat{\mathcal{N}}_\alpha ]   \hat{\mathcal{Q}}_\alpha (-\tau)\bigr> \nonumber \\ 
+ \bigl<  \hat{q}_\alpha \hat{q}_\alpha (-\tau) \bigr> ^* \bigl< [ \hat{\mathcal{Q}}_\alpha , \hat{\mathcal{N}}_\alpha ]   \hat{\mathcal{Q}}_\alpha (-\tau)\bigr> ^*  \Big\}.
\end{eqnarray}

Under the transformation of system operators in Eq. (\ref{eq:system_operators}) and (\ref{eq:system_operators_tau}), terms as $\langle \hat{q}_\alpha \hat{q}_\alpha (-\tau) \rangle$ are reduces to second moments of normal mode operators. Due to the partial secular approximation, highly oscillating contributions coming from terms as $\langle \hat{c}_i \hat{c}_j \rangle$ whit $\{ i,j \}= \{ 1,2 \}$ are vanished. In the strong coupling regimen, this approximation reduces the system terms in Eq. (\ref{eq:flux_2}) to
\begin{widetext}
\begin{eqnarray}   
\langle \hat{q}_a \hat{q}_a (-\tau) \rangle = e^{i\Omega_1 \tau} &\Big[ \cos^2(A) \langle \hat{c}_1^\dagger \hat{c}_1\rangle   - \cos(A) \sin(A)  \langle\hat{c}_1 \hat{c}_2^\dagger\rangle \Big]+ e^{-i\Omega_1 \tau} \Big[ \cos^2(A)  \langle\hat{c}_1 \hat{c}_1 ^\dagger\rangle  - \cos(A) \sin(A) \langle\hat{c}_1^\dagger  \hat{c}_2\rangle  \Big]\nonumber \\
+e^{i\Omega_2 \tau} &\Big[ \sin^2(A) \langle \hat{c}_2^\dagger \hat{c}_2\rangle  - \cos(A) \sin(A) \langle \hat{c}_1^\dagger \hat{c}_2 \rangle \Big] + e^{-i\Omega_2 \tau}  \Big[ \sin^2(A)  \langle \hat{c}_2 \hat{c}_2 ^\dagger\rangle  - \cos(A) \sin(A) \langle \hat{c}_1  \hat{c}_2^\dagger\rangle   \Big], \label{eq:system_CG_a} \\
\langle \hat{q}_b \hat{q}_b (-\tau) \rangle = e^{i\Omega_2 \tau} &\Big[ \cos^2(A) \langle \hat{c}_2^\dagger \hat{c}_2 \rangle   + \cos(A) \sin(A)  \langle\hat{c}_1^\dagger \hat{c}_2\rangle  \Big] + e^{-i\Omega_2 \tau}  \Big[ \cos^2(A)  \langle\hat{c}_2 \hat{c}_2 ^\dagger\rangle  + \cos(A) \sin(A)  \langle\hat{c}_1  \hat{c}_2^\dagger \rangle   \Big] \nonumber \\
+e^{i\Omega_1 \tau} &\Big[ \sin^2(A) \langle \hat{c}_1^\dagger \hat{c}_1 \rangle  + \cos(A) \sin(A) \langle \hat{c}_1 \hat{c}_2^\dagger \rangle \Big] + e^{-i\Omega_1 \tau}  \Big[ \sin^2(A) \langle \hat{c}_1 \hat{c}_1 ^\dagger\rangle   + \cos(A) \sin(A) \langle \hat{c}_1 ^\dagger \hat{c}_2 \rangle  \Big]. \label{eq:system_CG_b} 
\end{eqnarray}
\end{widetext}
On the other hand, considering the thermal state of the baths, bath terms in Eq. (\ref{eq:flux_2}) gives
\begin{eqnarray}\label{eq:third_bath}
\bigl< [ \hat{\mathcal{Q}}_\alpha , \hat{\mathcal{N}}_\alpha ]   \hat{\mathcal{Q}}_\alpha (-\tau)\bigr> =\gamma_\alpha\int_{0}^{\infty}  {\rm d} \omega \mathcal{J}(\omega)   \cos(\omega \tau) &&\nonumber \\
-i\gamma_\alpha\int_{0}^{\infty}  {\rm d} \omega \mathcal{J}(\omega)  \coth\left( \frac{\beta\omega}{2} \right) \sin(\omega \tau).&&
\end{eqnarray}
Previous terms in Eq. (\ref{eq:third_bath}) are such that
\begin{equation}\label{eq:third_correlation}
\int_0^{\infty} {\rm d}\tau \langle [ \hat{\mathcal{Q}}_\alpha , \hat{\mathcal{N}}_\alpha ]   \hat{\mathcal{Q}}_\alpha (-\tau)\rangle e^{\pm i \omega \tau} = \pm \gamma_\alpha\mathcal{C}( \pm \omega).
\end{equation}
whit $\mathcal{C}( \omega)$in Eq. (\ref{eq:def_bath_correlations}). Finally, expressions and relations in Eq. (\ref{eq:system_CG_a}-\ref{eq:third_correlation}), reduces Eq. (\ref{eq:flux_2}) to the final expressions in Eq. (\ref{eq:bath_particles_a}) and (\ref{eq:bath_particles_b}) for each bath.

\section{Prove of the non-Gibbs steady state}\label{appendix:non-canonical state} 

Here we derive the result of evaluating the Gibbs state in the system dynamics, i.e., 
\begin{widetext}
\begin{eqnarray}\label{eq:def_non-Gibbs_state}
\frac{\rm d}{\rm dt} \hat{\rho}_S \Bigg|_{\hat{\rho}_S= \hat{\rho}_{G}} &=& \sum_i \Big\{-i \overline{\Omega}_i \left[  \hat{c}_i^\dagger \hat{c}_i ,  \hat{\rho}_G  \right] +   \Gamma_i (\Omega_i) \mathcal{L}_{\hat{c}_i} [\hat{\rho}_G]  +\Gamma_i (-\Omega_i) \mathcal{L}_{\hat{c}_i^\dagger} [\hat{\rho}_G] \Big\}   \nonumber \\
&+& \sum_{i,j\neq i} \Big\{-i h_{i,j}^{(\rm s)} \left[  \hat{c}_i^\dagger \hat{c}_j , \hat{\rho}_G \right] +    \Gamma_{i,j}^{(\rm s)}(\Omega_i,\Omega_j) \mathcal{L}_{\hat{c}_i^\dagger, \hat{c}_j} [\hat{\rho}_G] 
+ \Gamma_{i,j}^{(\rm s)}(-\Omega_j, - \Omega_i)^* \mathcal{L}_{\hat{c}_i, \hat{c}_j^\dagger} [\hat{\rho}_G]  \Big\}, 
\end{eqnarray}
\end{widetext}
to demonstrate that the system in general relaxes in a non-Gibbs state. We extend the analysis beyond the strong coupling regime by considering the system dynamics in Eq. (\ref{eq:full_Redfield}) in the partial secular form.

By expressing the system Hamiltonian in diagonal form (see Eq. (\ref{eq:diagonal_Hamiltonian})), the Gibbs state in Eq. (\ref{eq:canonical_state}) satisfies the property $\hat{\rho}_{G} \hat{c}_i = \exp (\beta \omega_i ) \hat{c}_i \hat{\rho}_{G}$, reducing terms from Eq. (\ref{eq:def_non-Gibbs_state}) to
\begin{eqnarray}\label{eq:canonical_terms}
    &&\left[ \hat{c}_i^\dagger \hat{c}_j , \hat{\rho}_G   \right] =(1-\delta_{i,j})\left[ \frac{1}{n^{\rm th}(\omega_i)} - \frac{1}{n^{\rm th}(\omega_j)}  \right]  \hat{c}_j \hat{\rho}_G \hat{c}_i^\dagger,\nonumber \label{eq:non_Gibbs_1} \\
    \\
    &&\mathcal{L}_{\hat{c}_i}[\hat{\rho}_G] = e^{- \beta \omega_i } \hat{\rho}_G - \frac{e^{- \beta\omega_i }}{n^{\rm th}(\omega_i)} \hat{c}_i^\dagger \hat{c}_i \hat{\rho}_G ,\label{eq:non_Gibbs_2} \\
    &&\mathcal{L}_{\hat{c}_i^\dagger}[\hat{\rho}_G] = - \hat{\rho}_G + \frac{1}{n^{\rm th}(\omega_i)} \hat{c}_i^\dagger \hat{c}_i \hat{\rho}_G , \label{eq:non_Gibbs_3}\\
    &&\mathcal{L}_{\hat{c}_i , \hat{c}_j^\dagger}[\hat{\rho}_G] =  \frac{1}{2} \hat{c}_i  \hat{\rho}_G  \hat{c}_j^\dagger\Bigg\{ \frac{e^{\beta \omega_i}}{n^{\rm th}(\omega_j)} + \frac{e^{\beta \omega_j}}{n^{\rm th} (\omega_i)} \Bigg\}, \label{eq:non_Gibbs_4}\\
    &&\mathcal{L}_{\hat{c}_i ^\dagger , \hat{c}_j }[\hat{\rho}_G] =  - \frac{1}{2}  \hat{c}_j \hat{\rho}_G  \hat{c}_i^\dagger\Bigg\{ \frac{1}{n^{\rm th}(\omega_i)} + \frac{1}{n^{\rm th} (\omega_j)} \Bigg\} \nonumber, \label{eq:non_Gibbs_5}\\
\end{eqnarray}
for $\{ i,j \} = \{ 1,2\}$ the normal mode indexes. Using Eq. (\ref{eq:non_Gibbs_1}-\ref{eq:non_Gibbs_5}), and writing the master equation parameters in terms of the real and imaginary part of $\mathcal{C}(\omega)$, Eq. (\ref{eq:def_non-Gibbs_state}) is reduced to
\begin{eqnarray}\label{eq:non-canonical_CG}
\frac{\rm d}{\rm dt} \hat{\rho}_S \Bigg|_{\hat{\rho}_S = \hat{\rho}_G} = i\sum_{\alpha=\{ a,b \}} \gamma_\alpha  \left[ \frac{f(\omega_1)}{n^{\rm th}(\omega_2)} - \frac{f(\omega_2)}{n^{\rm th}(\omega_1)} \right]&& \nonumber\\
\times  \left( \chi_{\alpha,1}^* \chi_{\alpha,2} \hat{c}_2 \hat{\rho}_G \hat{c}_1^\dagger  -  {\rm h.c.} \right)&&.
\end{eqnarray}
where $\chi_{\alpha,i} = U_{\alpha,i} + V_{\alpha,i}^*$ are transformation factors. In Eq. (\ref{eq:non-canonical_CG}), detail balance condition vanishes the contributions from $\Gamma(\omega)$. However, non-zero contributions from non-secular terms are absorbed in the functions $f(\omega) = \mathcal{S}(\omega) + \exp(\beta \omega) \mathcal{S}(-\omega)$. For strongly interacting oscillators, the factors $\{ \chi_{a,1},\chi_{a,2},\chi_{b,1},\chi_{b,2} \} = \{ \cos(A),-\sin(A),\sin(A),\cos(A) \}$ reduces Eq. (\ref{eq:non-canonical_CG}) to Eq. (\ref{eq:non-canonical}).

\section{Steady oscillator occupations}\label{appendix:oscillator occupation}
Here we obtain expressions for steady occupation of the strongly interacting oscillators. From the system dynamics in Eq. (\ref{eq:CG_Redfield}), we construct dynamics of the normal mode observables,
\begin{equation}
    \bm{X}  = \Big[ \langle \hat{c}_1 ^\dagger \hat{c}_1 \rangle , \langle \hat{c}_2 ^\dagger \hat{c}_2 \rangle , {\rm Re}\langle \hat{c}_1  \hat{c}_2^\dagger\rangle , {\rm Im} \langle \hat{c}_1 \hat{c}_2^\dagger \rangle \Big]^{\rm T},
\end{equation}
as a vectorial equation 
\begin{equation}\label{eq:observable_dynamics}
\frac{\rm d}{\rm dt}  \bm{X}  = \mathcal{M}   \bm{X} + \bm{c},
\end{equation}
where
\begin{eqnarray}
&\mathcal{M} = \begin{pmatrix}
    -\gamma_1     & 0 & -\mathcal{R}_1  & -\mathcal{I}_1 \\
   0       & -\gamma_2 & \mathcal{R}_2  & \mathcal{I}_2 \\
    \mathcal{R}_2/2      & -\mathcal{R}_1/2 & -\overline{\gamma}  & (\overline{\Omega}_1 - \overline{\Omega}_2) \\
    \mathcal{I}_2/2     & -\mathcal{I}_1/2 & -(\overline{\Omega}_1 - \overline{\Omega}_2)  & -\overline{\gamma}
\end{pmatrix},\\
&\bm{c} = \Big[  \gamma_1n^{\rm th}(\Omega_1) ,  \gamma_2n^{\rm th}(\Omega_2), {\rm Re} \{\Gamma_{21} \}, {\rm Im}\{\Gamma_{21}\} \Big]^{\rm T}.
\end{eqnarray}
In Eq. (\ref{eq:observable_dynamics}), the secular contributions describe that normal mode occupations and coherence evolves independently, being represented by the secular master equation parameters 
\begin{eqnarray}
    &\gamma_i = \Gamma_i(\Omega_i) - \Gamma_i(-\Omega_i),\hspace{0.8cm}\overline{\gamma} = [\gamma_1 + \gamma_2]/2,
\end{eqnarray}
whit $i=\{ 1,2 \}$ the normal mode index. Meanwhile, the non-secular contributions in Eq. (\ref{eq:observable_dynamics}),  couple the evolution of normal mode occupations and coherence, being represented by the non-secular master equation parameters
\begin{eqnarray}
{\mathcal R}_i &=& 2{\rm Im} \overline{g} -(-1)^i{\rm Re} \left\{ \Gamma_{12} - \Gamma_{21}^* \right\}, \label{eq:def_R_i}\\
{\mathcal I}_i &=& 2{\rm Re} \overline{g} +(-1)^i {\rm Im} \left\{ \Gamma_{12} - \Gamma_{21}^* \right\}. \label{eq:def_I_i}
\end{eqnarray}

In the steady state, normal modes observables, $\langle \hat{c}_1^\dagger \hat{c}_1 \rangle_{\rm ss}$, $\langle \hat{c}_2^\dagger \hat{c}_2 \rangle_{\rm ss}$ and  $\langle \hat{c}_1 \hat{c}_2^\dagger \rangle_{\rm ss}$, can be computed by setting to zero the left side of Eq. (\ref{eq:observable_dynamics}). Let us write the steady coherence in terms of steady occupations as
\begin{eqnarray}
{\rm Re} \langle \hat{c}_1 \hat{c}_2^\dagger \rangle_{\rm ss} &=&  {\rm Re} \Big\{ \mathcal{Y} \Gamma_{21}^* \Big\} +\frac{1}{2}{\rm Re} \Big\{ \mathcal{Y} \mathcal{Z}_2^* \Big\} \langle \hat{c}_1^\dagger \hat{c}_1 \rangle_{\rm ss} \nonumber\\
 &-& \frac{1}{2}{\rm Re} \Big\{ \mathcal{Y} \mathcal{Z}_1^* \Big\} \langle \hat{c}_2^\dagger \hat{c}_2 \rangle_{\rm ss}   ,\label{eq:def_real_normal_mode_coherence}\\
{\rm Im} \langle \hat{c}_1 \hat{c}_2^\dagger \rangle_{\rm ss} &=&  - {\rm Im} \Big\{ \mathcal{Y} \Gamma_{21}^* \Big\} -\frac{1}{2}{\rm Im} \Big\{ \mathcal{Y} \mathcal{Z}_2^* \Big\} \langle \hat{c}_1^\dagger \hat{c}_1 \rangle_{\rm ss} \nonumber \\
&+& \frac{1}{2}{\rm Im} \Big\{ \mathcal{Y} \mathcal{Z}_1^* \Big\} \langle \hat{c}_2^\dagger \hat{c}_2 \rangle_{\rm ss}  ,\label{eq:def_imaginary_normal_mode_coherence}
\end{eqnarray}
where 
\begin{eqnarray}
\mathcal{Y} = \frac{\overline{\gamma} + i (\overline{\Omega}_1 - \overline{\Omega}_2)}{\overline{\gamma}^2 + (\overline{\Omega}_1 - \overline{\Omega}_2)^2},\hspace{0.8cm}\mathcal{Z}_j = \mathcal{R}_j + i \mathcal{I}_j. \label{eq:def_Y-Z}
\end{eqnarray}
Equations (\ref{eq:def_real_normal_mode_coherence}) and (\ref{eq:def_imaginary_normal_mode_coherence}) reduce normal mode occupations to
\begin{eqnarray}
    \langle \hat{c}_1^\dagger \hat{c}_1 \rangle_{\rm ss} &=& \Big\{ n_1 + \frac{\delta_{11}}{1 + \delta_{21}}  n_2 \Big\} \Big\{ 1 + \delta_{12} - \frac{\delta_{11} \delta_{22}}{1+\delta_{21} } \Big\}^{-1},\nonumber \\\label{eq:normal_mode_occupations_1}  \\  
    \langle \hat{c}_2^\dagger \hat{c}_2 \rangle_{\rm ss} &=&  \Big\{ n_2 + \frac{\delta_{22}}{1 + \delta_{12}} n_1 \Big\}  \Big\{ 1 + \delta_{21} - \frac{\delta_{11} \delta_{22}}{1+\delta_{12} } \Big\}^{-1}.\nonumber \\ \label{eq:normal_mode_occupations_2} \\ \nonumber  
\end{eqnarray}
where
\begin{eqnarray}
n_i &=& n^{\rm th}(\Omega_i) +\frac{(-1)^i}{\gamma_i }{\rm Re} \Big\{ \mathcal{Y}\mathcal{Z}_i \Gamma_{21}^* \Big\},\label{eq:cross_parameters_1_n} \\
\delta_{ij} &=& \frac{1}{2\gamma_i }{\rm Re} \Big\{ \mathcal{Y} \mathcal{Z}_i \mathcal{Z}_j^* \Big\}, 
 \label{eq:cross_parameters_2_t}
\end{eqnarray}
for $\{ i,j \} = \{ 1,2 \}$. In the secular approximation, i.e., setting $\{ \overline{g}, \Gamma_{12},\Gamma_{21} \}=0$, Eq.  (\ref{eq:normal_mode_occupations_1}) and (\ref{eq:normal_mode_occupations_2}) lead to steady thermal occupations, $\langle \hat{c}_1^\dagger \hat{c}_1 \rangle_{\rm ss} = n^{\rm th}(\Omega_1)$ and $\langle \hat{c}_2^\dagger \hat{c}_2 \rangle_{\rm ss} = n^{\rm th}(\Omega_2)$, while Eq. (\ref{eq:def_real_normal_mode_coherence}) and (\ref{eq:def_imaginary_normal_mode_coherence}) lead to zero coherence, which are values consistent with the Gibbs steady state of the system.

\subsection{Approximated steady occupations of resonant oscillators}\label{appendix:compensation_relation}

Based on the exact expressions for the normal mode occupations and coherence in Appendix \ref{appendix:oscillator occupation}, here we derive approximated expressions for the steady occupation of oscillators A and B under the resonant condition $\Delta=0$ and strong coupling. In this regime, normal mode frequencies in Eq. (\ref{eq:omega_1}) and (\ref{eq:omega_2}) are $\Omega_1 = \omega_b + g$ and $\Omega_2 = \omega_b -g$, respectively, and x$\cos(A)=1/\sqrt{2}=\sin(A)$. The function $\mathcal{J}(\omega)$ in Eq. (\ref{eq:dimensionless_spectral_density}) evaluated at the normal mode frequencies is
\begin{eqnarray}
\mathcal{J}(\Omega_i) \approx \frac{1}{2\pi}\left[ 1 -(-1)^i \frac{g}{\omega_b} \right], \label{eq:approx_spectral_densities_1}\end{eqnarray}
up to first order in $g/\omega_b \ll 1$ whit $i=\{ 1,2 \}$. Based on the last expressions, the parameters of the master equations in Eq. (\ref{eq:CG_Redfield}) are approximated to
\begin{eqnarray}
& &\gamma_i\approx  \frac{\gamma_a + \gamma_b}{2}  \left[ 1 -(-1)^i \frac{g}{\omega_b} \right] ,\label{eq:approx_CG_parameters_gamma_i}\\
& &{\rm Im}\{ \overline{g}\} \approx   \left[  \frac{\gamma_b - \gamma_a}{4}\right] \frac{g}{\omega_b},\label{eq:approx_CG_parameters_g_12}\\
& &{\rm Re} \{ \Gamma_{12} \}  \approx   \left[  \frac{\gamma_b - \gamma_a}{4}\right] \label{eq:approx_CG_parameters_eta_12}\\
&  & \times   \Big\{ 2+ n^{\rm th}(\Omega_1)  +  n^{\rm th} (\Omega_2) + \frac{g}{\omega_b} \left[ n^{\rm th}(\Omega_1) - n^{\rm th}(\Omega_2) \right]  \Big\},\nonumber\\
& &{\rm Re} \{ \Gamma_{21} \}  \approx  \left[  \frac{\gamma_b - \gamma_a}{4}\right] \label{eq:approx_CG_parameters_eta_21} \\
& & \times   \Big\{ n^{\rm th}(\Omega_1)  +   n^{\rm th} (\Omega_2) + \frac{g}{\omega_b} \left[ n^{\rm th}(\Omega_1) - n^{\rm th}(\Omega_2) \right]  \Big\} \nonumber.
\end{eqnarray}
In Eq. (\ref{eq:approx_CG_parameters_gamma_i} - \ref{eq:approx_CG_parameters_eta_21}), we approximate terms that depend on the real part of $\mathcal{C}(\omega)$. Meanwhile terms that depends on the imaginary part of $\mathcal{C}(\omega)$ are computed numerically in the strong coupling regime, obtaining that $|{\rm Re}\{ \overline{g} \}| \approx 0$ $ |{\rm Im} \{ \Gamma_{12}  - \Gamma_{21}^*\}| \approx 0$. All these approximations reduces the non-secular parameters in Eq. (\ref{eq:def_R_i})  and (\ref{eq:def_I_i}) to 
\begin{eqnarray}\label{eq:approx_R_i-I_i}
    \mathcal{R}_1  = -\mathcal{R}_2 \approx \frac{\gamma_b - \gamma_a}{2}, \hspace{0.8cm}
    \mathcal{I}_1 = \mathcal{I}_2 \approx 0,
\end{eqnarray}
for the lowest order of $g/\omega_b$. 

Considering negligible Lamb-shift of the normal mode frequencies ($\overline{\Omega}_i\approx \Omega_i$), the strong coupling regime, manifested in $\overline{\Omega}_1 - \overline{\Omega}_2 \approx 2g \gg \overline{\gamma} \approx (\gamma_1 + \gamma_2)/2$, and the relations in Eq. (\ref{eq:approx_R_i-I_i}) leads the terms in Eq. (\ref{eq:def_Y-Z}) to
\begin{eqnarray}
\mathcal{Y}\approx \frac{(\gamma_a + \gamma_b)}{8g^2}+ \frac{i}{2g},\hspace{0.8cm} \mathcal{Z}_j \approx (-1)^j \frac{(\gamma_a - \gamma_b)}{2}.
\end{eqnarray}
The previous approximations lead  to $\delta_{i,j} \approx 0$ (defined in Eq. (\ref{eq:cross_parameters_2_t})). Therefore, normal mode oscillator occupations in Eq. (\ref{eq:normal_mode_occupations_1}) and (\ref{eq:normal_mode_occupations_2}) are approximated to $\langle \hat{c}_1^\dagger \hat{c}_1 \rangle_{\rm ss}\approx n_1$ and $\langle \hat{c}_2^\dagger \hat{c}_2 \rangle_{\rm ss}\approx n_2$. Finally, using $n_i$ in Eq. (\ref{eq:cross_parameters_1_n}), normal mode occupations are reduced to
\begin{eqnarray}
 \langle \hat{c}_i^\dagger \hat{c}_i \rangle_{\rm ss} &-& n^{\rm th}(\Omega_i) \approx \\
 &+& \left[ 1+ (-1)^i \left(\frac{g}{\omega_b} \right) \right]\frac{ (\gamma_a - \gamma_b ) }{2g(\gamma_a + \gamma_b)} {\rm Im} \{ \Gamma_{21} \} .\nonumber \label{eq:approx_normal_modes}\end{eqnarray}
The same scheme of approximations applied for obtaining oscillator occupations in Eq. (\ref{eq:approx_normal_modes}) gives an steady normal mode coherence (see Eq. (\ref{eq:def_real_normal_mode_coherence}) and (\ref{eq:def_imaginary_normal_mode_coherence}))
\begin{eqnarray}
{\rm Re} \langle \hat{c}_1 \hat{c}_2 ^\dagger \rangle_{\rm ss} & \approx & \frac{1}{2g} {\rm Im} \{ \Gamma_{21} \},\label{eq:approx_real_coherence}\\
{\rm Im} \langle \hat{c}_1 \hat{c}_2 ^\dagger \rangle_{\rm ss} & \approx &  \frac{\gamma_a \gamma_b}{2g^2 (\gamma_a + \gamma_b) } {\rm Im} \{ \Gamma_{21} \}. \label{eq:approx_imaginary_coherence}
\end{eqnarray}

Under resonant condition, the steady occupation of oscillators A and B are related to the normal modes observables in Eq. (\ref{eq:approx_normal_modes}-\ref{eq:approx_imaginary_coherence}) according to the Bogoliubov transformations
\begin{eqnarray}
\langle \hat{a}^\dagger \hat{a} \rangle_{\rm ss} &=& \frac{1}{2} \langle \hat{c}_1^\dagger \hat{c}_1 \rangle_{\rm ss} + \frac{1}{2} \langle \hat{c}_2^\dagger \hat{c}_2 \rangle_{\rm ss} - {\rm Re} \langle \hat{c}_1 \hat{c}_2^\dagger\rangle_{\rm ss} \label{eq:approx_oscillator_occupation_a}\\
&\approx & \frac{1}{2} \left[ n^{\rm th}(\Omega_1) + n^{\rm th}(\Omega_2) \right] - \frac{\gamma_b }{g(\gamma_a + \gamma_b)}{\rm Im} \{ \Gamma_{21} \},\nonumber \\
\langle \hat{b}^\dagger \hat{b} \rangle_{\rm ss} &= &\frac{1}{2} \langle \hat{c}_1^\dagger \hat{c}_1 \rangle_{\rm ss} + \frac{1}{2} \langle \hat{c}_2^\dagger \hat{c}_2 \rangle_{\rm ss} + {\rm Re} \langle \hat{c}_1 \hat{c}_2^\dagger\rangle_{\rm ss} \label{eq:approx_oscillator_occupation_b}\\
&\approx & \frac{1}{2} \left[ n^{\rm th}(\Omega_1) + n^{\rm th}(\Omega_2) \right] + \frac{\gamma_a }{g(\gamma_a + \gamma_b)}{\rm Im} \{ \Gamma_{21} \}.\nonumber
\end{eqnarray}
In the temperature regime regime $T \gtrsim  g$, the condition $ n^{\rm th}(\Omega_1) + n^{\rm th}(\Omega_2) \approx 2 n^{\rm th}(\omega_b)$ reduces Eq. (\ref{eq:approx_oscillator_occupation_a}) and  (\ref{eq:approx_oscillator_occupation_b}) to the final expressions of steady oscillator occupations in Eq. (\ref{eq:approx_resonant_occupation_a}) and (\ref{eq:approx_resonant_occupation_b}). These expressions are written in  terms  of steady coherence ${\rm Im}  \langle \hat{a} \hat{b}^\dagger \rangle_{\rm ss} = {\rm Im} \langle \hat{c}_1 \hat{c}_2^\dagger \rangle_{\rm ss}$, which was derived in Eq. (\ref{eq:approx_imaginary_coherence}).

%

\bibliographystyle{apsrev4-1}
\bibliography{noncanonical}

\providecommand{\noopsort}[1]{}\providecommand{\singleletter}[1]{#1}%
\begin{thebibliography}{49}%
\makeatletter
\providecommand \@ifxundefined [1]{%
 \@ifx{#1\undefined}
}%
\providecommand \@ifnum [1]{%
 \ifnum #1\expandafter \@firstoftwo
 \else \expandafter \@secondoftwo
 \fi
}%
\providecommand \@ifx [1]{%
 \ifx #1\expandafter \@firstoftwo
 \else \expandafter \@secondoftwo
 \fi
}%
\providecommand \natexlab [1]{#1}%
\providecommand \enquote  [1]{``#1''}%
\providecommand \bibnamefont  [1]{#1}%
\providecommand \bibfnamefont [1]{#1}%
\providecommand \citenamefont [1]{#1}%
\providecommand \href@noop [0]{\@secondoftwo}%
\providecommand \href [0]{\begingroup \@sanitize@url \@href}%
\providecommand \@href[1]{\@@startlink{#1}\@@href}%
\providecommand \@@href[1]{\endgroup#1\@@endlink}%
\providecommand \@sanitize@url [0]{\catcode `\\12\catcode `\$12\catcode
  `\&12\catcode `\#12\catcode `\^12\catcode `\_12\catcode `\%12\relax}%
\providecommand \@@startlink[1]{}%
\providecommand \@@endlink[0]{}%
\providecommand \url  [0]{\begingroup\@sanitize@url \@url }%
\providecommand \@url [1]{\endgroup\@href {#1}{\urlprefix }}%
\providecommand \urlprefix  [0]{URL }%
\providecommand \Eprint [0]{\href }%
\providecommand \doibase [0]{http://dx.doi.org/}%
\providecommand \selectlanguage [0]{\@gobble}%
\providecommand \bibinfo  [0]{\@secondoftwo}%
\providecommand \bibfield  [0]{\@secondoftwo}%
\providecommand \translation [1]{[#1]}%
\providecommand \BibitemOpen [0]{}%
\providecommand \bibitemStop [0]{}%
\providecommand \bibitemNoStop [0]{.\EOS\space}%
\providecommand \EOS [0]{\spacefactor3000\relax}%
\providecommand \BibitemShut  [1]{\csname bibitem#1\endcsname}%
\let\auto@bib@innerbib\@empty
\bibitem [{\citenamefont {Eastham}\ \emph {et~al.}(2016)\citenamefont
  {Eastham}, \citenamefont {Kirton}, \citenamefont {Cammack}, \citenamefont
  {Lovett},\ and\ \citenamefont {Keeling}}]{eastham2016bath}%
  \BibitemOpen
  \bibfield  {author} {\bibinfo {author} {\bibfnamefont {P.~R.}\ \bibnamefont
  {Eastham}}, \bibinfo {author} {\bibfnamefont {P.}~\bibnamefont {Kirton}},
  \bibinfo {author} {\bibfnamefont {H.~M.}\ \bibnamefont {Cammack}}, \bibinfo
  {author} {\bibfnamefont {B.~W.}\ \bibnamefont {Lovett}}, \ and\ \bibinfo
  {author} {\bibfnamefont {J.}~\bibnamefont {Keeling}},\ }\href {\doibase
  10.1103/PhysRevA.94.012110} {\bibfield  {journal} {\bibinfo  {journal} {Phys.
  Rev. A}\ }\textbf {\bibinfo {volume} {94}},\ \bibinfo {pages} {012110}
  (\bibinfo {year} {2016})}\BibitemShut {NoStop}%
\bibitem [{\citenamefont {Sousa}\ \emph {et~al.}(2022)\citenamefont {Sousa},
  \citenamefont {Vieira}, \citenamefont {Santos},\ and\ \citenamefont
  {da~Paz}}]{Sousa}%
  \BibitemOpen
  \bibfield  {author} {\bibinfo {author} {\bibfnamefont {J.~F.}\ \bibnamefont
  {Sousa}}, \bibinfo {author} {\bibfnamefont {C.~H.~S.}\ \bibnamefont
  {Vieira}}, \bibinfo {author} {\bibfnamefont {J.~F.~G.}\ \bibnamefont
  {Santos}}, \ and\ \bibinfo {author} {\bibfnamefont {I.~G.}\ \bibnamefont
  {da~Paz}},\ }\href {\doibase 10.1103/PhysRevA.106.032401} {\bibfield
  {journal} {\bibinfo  {journal} {Phys. Rev. A}\ }\textbf {\bibinfo {volume}
  {106}},\ \bibinfo {pages} {032401} (\bibinfo {year} {2022})}\BibitemShut
  {NoStop}%
\bibitem [{\citenamefont {Purkayastha}\ \emph {et~al.}(2020)\citenamefont
  {Purkayastha}, \citenamefont {Guarnieri}, \citenamefont {Mitchison},
  \citenamefont {Filip},\ and\ \citenamefont {Goold}}]{purkayastha2020tunable}%
  \BibitemOpen
  \bibfield  {author} {\bibinfo {author} {\bibfnamefont {A.}~\bibnamefont
  {Purkayastha}}, \bibinfo {author} {\bibfnamefont {G.}~\bibnamefont
  {Guarnieri}}, \bibinfo {author} {\bibfnamefont {M.~T.}\ \bibnamefont
  {Mitchison}}, \bibinfo {author} {\bibfnamefont {R.}~\bibnamefont {Filip}}, \
  and\ \bibinfo {author} {\bibfnamefont {J.}~\bibnamefont {Goold}},\ }\href
  {\doibase 10.1038/s41534-020-0256-6} {\bibfield  {journal} {\bibinfo
  {journal} {npj Quantum Information}\ }\textbf {\bibinfo {volume} {6}},\
  \bibinfo {pages} {27} (\bibinfo {year} {2020})}\BibitemShut {NoStop}%
\bibitem [{\citenamefont {Pires}\ \emph {et~al.}(2018)\citenamefont {Pires},
  \citenamefont {Silva}, \citenamefont {deAzevedo}, \citenamefont
  {Soares-Pinto},\ and\ \citenamefont {Filgueiras}}]{pires2018coherence}%
  \BibitemOpen
  \bibfield  {author} {\bibinfo {author} {\bibfnamefont {D.~P.}\ \bibnamefont
  {Pires}}, \bibinfo {author} {\bibfnamefont {I.~A.}\ \bibnamefont {Silva}},
  \bibinfo {author} {\bibfnamefont {E.~R.}\ \bibnamefont {deAzevedo}}, \bibinfo
  {author} {\bibfnamefont {D.~O.}\ \bibnamefont {Soares-Pinto}}, \ and\
  \bibinfo {author} {\bibfnamefont {J.~G.}\ \bibnamefont {Filgueiras}},\ }\href
  {\doibase 10.1103/PhysRevA.98.032101} {\bibfield  {journal} {\bibinfo
  {journal} {Phys. Rev. A}\ }\textbf {\bibinfo {volume} {98}},\ \bibinfo
  {pages} {032101} (\bibinfo {year} {2018})}\BibitemShut {NoStop}%
\bibitem [{\citenamefont {Joo}\ \emph {et~al.}(2011)\citenamefont {Joo},
  \citenamefont {Munro},\ and\ \citenamefont {Spiller}}]{joo2011quantum}%
  \BibitemOpen
  \bibfield  {author} {\bibinfo {author} {\bibfnamefont {J.}~\bibnamefont
  {Joo}}, \bibinfo {author} {\bibfnamefont {W.~J.}\ \bibnamefont {Munro}}, \
  and\ \bibinfo {author} {\bibfnamefont {T.~P.}\ \bibnamefont {Spiller}},\
  }\href {\doibase 10.1103/PhysRevLett.107.083601} {\bibfield  {journal}
  {\bibinfo  {journal} {Phys. Rev. Lett.}\ }\textbf {\bibinfo {volume} {107}},\
  \bibinfo {pages} {083601} (\bibinfo {year} {2011})}\BibitemShut {NoStop}%
\bibitem [{\citenamefont {Albash}\ and\ \citenamefont
  {Lidar}(2015)}]{albash2015decoherence}%
  \BibitemOpen
  \bibfield  {author} {\bibinfo {author} {\bibfnamefont {T.}~\bibnamefont
  {Albash}}\ and\ \bibinfo {author} {\bibfnamefont {D.~A.}\ \bibnamefont
  {Lidar}},\ }\href {\doibase 10.1103/PhysRevA.91.062320} {\bibfield  {journal}
  {\bibinfo  {journal} {Phys. Rev. A}\ }\textbf {\bibinfo {volume} {91}},\
  \bibinfo {pages} {062320} (\bibinfo {year} {2015})}\BibitemShut {NoStop}%
\bibitem [{\citenamefont {Albash}\ \emph {et~al.}(2012)\citenamefont {Albash},
  \citenamefont {Boixo}, \citenamefont {Lidar},\ and\ \citenamefont
  {Zanardi}}]{albash2012quantum}%
  \BibitemOpen
  \bibfield  {author} {\bibinfo {author} {\bibfnamefont {T.}~\bibnamefont
  {Albash}}, \bibinfo {author} {\bibfnamefont {S.}~\bibnamefont {Boixo}},
  \bibinfo {author} {\bibfnamefont {D.~A.}\ \bibnamefont {Lidar}}, \ and\
  \bibinfo {author} {\bibfnamefont {P.}~\bibnamefont {Zanardi}},\ }\href
  {\doibase 10.1088/1367-2630/14/12/123016} {\bibfield  {journal} {\bibinfo
  {journal} {New Journal of Physics}\ }\textbf {\bibinfo {volume} {14}},\
  \bibinfo {pages} {123016} (\bibinfo {year} {2012})}\BibitemShut {NoStop}%
\bibitem [{\citenamefont {Trushechkin}\ \emph {et~al.}(2022)\citenamefont
  {Trushechkin}, \citenamefont {Merkli}, \citenamefont {Cresser},\ and\
  \citenamefont {Anders}}]{trushechkin2022open}%
  \BibitemOpen
  \bibfield  {author} {\bibinfo {author} {\bibfnamefont {A.~S.}\ \bibnamefont
  {Trushechkin}}, \bibinfo {author} {\bibfnamefont {M.}~\bibnamefont {Merkli}},
  \bibinfo {author} {\bibfnamefont {J.~D.}\ \bibnamefont {Cresser}}, \ and\
  \bibinfo {author} {\bibfnamefont {J.}~\bibnamefont {Anders}},\ }\href
  {\doibase 10.1116/5.0073853} {\bibfield  {journal} {\bibinfo  {journal} {AVS
  Quantum Science}\ }\textbf {\bibinfo {volume} {4}},\ \bibinfo {pages}
  {012301} (\bibinfo {year} {2022})}\BibitemShut {NoStop}%
\bibitem [{\citenamefont {Davidovi{\'{c}}}(2020)}]{davidovic2020completely}%
  \BibitemOpen
  \bibfield  {author} {\bibinfo {author} {\bibfnamefont {D.}~\bibnamefont
  {Davidovi{\'{c}}}},\ }\href {\doibase 10.22331/q-2020-09-21-326} {\bibfield
  {journal} {\bibinfo  {journal} {{Quantum}}\ }\textbf {\bibinfo {volume}
  {4}},\ \bibinfo {pages} {326} (\bibinfo {year} {2020})}\BibitemShut {NoStop}%
\bibitem [{\citenamefont {Ahn}\ \emph {et~al.}(2023)\citenamefont {Ahn},
  \citenamefont {Triana}, \citenamefont {Recabal}, \citenamefont {Herrera},\
  and\ \citenamefont {Simpkins}}]{ahn2023modification}%
  \BibitemOpen
  \bibfield  {author} {\bibinfo {author} {\bibfnamefont {W.}~\bibnamefont
  {Ahn}}, \bibinfo {author} {\bibfnamefont {J.~F.}\ \bibnamefont {Triana}},
  \bibinfo {author} {\bibfnamefont {F.}~\bibnamefont {Recabal}}, \bibinfo
  {author} {\bibfnamefont {F.}~\bibnamefont {Herrera}}, \ and\ \bibinfo
  {author} {\bibfnamefont {B.~S.}\ \bibnamefont {Simpkins}},\ }\href {\doibase
  10.1126/science.ade7147} {\bibfield  {journal} {\bibinfo  {journal}
  {Science}\ }\textbf {\bibinfo {volume} {380}},\ \bibinfo {pages} {1165}
  (\bibinfo {year} {2023})}\BibitemShut {NoStop}%
\bibitem [{\citenamefont {Roy}\ and\ \citenamefont
  {Hughes}(2011)}]{roy2011influence}%
  \BibitemOpen
  \bibfield  {author} {\bibinfo {author} {\bibfnamefont {C.}~\bibnamefont
  {Roy}}\ and\ \bibinfo {author} {\bibfnamefont {S.}~\bibnamefont {Hughes}},\
  }\href {\doibase 10.1103/PhysRevX.1.021009} {\bibfield  {journal} {\bibinfo
  {journal} {Phys. Rev. X}\ }\textbf {\bibinfo {volume} {1}},\ \bibinfo {pages}
  {021009} (\bibinfo {year} {2011})}\BibitemShut {NoStop}%
\bibitem [{\citenamefont {Palmieri}\ \emph {et~al.}(2009)\citenamefont
  {Palmieri}, \citenamefont {Abramavicius},\ and\ \citenamefont
  {Mukamel}}]{palmieri2009lindblad}%
  \BibitemOpen
  \bibfield  {author} {\bibinfo {author} {\bibfnamefont {B.}~\bibnamefont
  {Palmieri}}, \bibinfo {author} {\bibfnamefont {D.}~\bibnamefont
  {Abramavicius}}, \ and\ \bibinfo {author} {\bibfnamefont {S.}~\bibnamefont
  {Mukamel}},\ }\href {\doibase 10.1063/1.3142485} {\enquote {\bibinfo {title}
  {Lindblad equations for strongly coupled populations and coherences in
  photosynthetic complexes},}\ } (\bibinfo {year} {2009})\BibitemShut {NoStop}%
\bibitem [{\citenamefont {Herrera}\ and\ \citenamefont
  {Owrutsky}(2020)}]{Herrera2020}%
  \BibitemOpen
  \bibfield  {author} {\bibinfo {author} {\bibfnamefont {F.}~\bibnamefont
  {Herrera}}\ and\ \bibinfo {author} {\bibfnamefont {J.}~\bibnamefont
  {Owrutsky}},\ }\href {\doibase 10.1063/1.5136320} {\bibfield  {journal}
  {\bibinfo  {journal} {The Journal of Chemical Physics}\ }\textbf {\bibinfo
  {volume} {152}},\ \bibinfo {pages} {100902} (\bibinfo {year}
  {2020})}\BibitemShut {NoStop}%
\bibitem [{\citenamefont {Scali}\ \emph {et~al.}(2021)\citenamefont {Scali},
  \citenamefont {Anders},\ and\ \citenamefont {Correa}}]{scali2021local}%
  \BibitemOpen
  \bibfield  {author} {\bibinfo {author} {\bibfnamefont {S.}~\bibnamefont
  {Scali}}, \bibinfo {author} {\bibfnamefont {J.}~\bibnamefont {Anders}}, \
  and\ \bibinfo {author} {\bibfnamefont {L.~A.}\ \bibnamefont {Correa}},\
  }\href {\doibase 10.22331/q-2021-05-01-451} {\bibfield  {journal} {\bibinfo
  {journal} {{Quantum}}\ }\textbf {\bibinfo {volume} {5}},\ \bibinfo {pages}
  {451} (\bibinfo {year} {2021})}\BibitemShut {NoStop}%
\bibitem [{\citenamefont {Recabal}\ and\ \citenamefont
  {Herrera}(2026)}]{recabal2026}%
  \BibitemOpen
  \bibfield  {author} {\bibinfo {author} {\bibfnamefont {F.}~\bibnamefont
  {Recabal}}\ and\ \bibinfo {author} {\bibfnamefont {F.}~\bibnamefont
  {Herrera}},\ }\href {\doibase 10.1063/5.0319363} {\bibfield  {journal}
  {\bibinfo  {journal} {The Journal of Chemical Physics}\ }\textbf {\bibinfo
  {volume} {164}},\ \bibinfo {pages} {114113} (\bibinfo {year}
  {2026})}\BibitemShut {NoStop}%
\bibitem [{\citenamefont {Orgiu}\ \emph {et~al.}(2015)\citenamefont {Orgiu},
  \citenamefont {George}, \citenamefont {Hutchison}, \citenamefont {Devaux},
  \citenamefont {Dayen}, \citenamefont {Doudin}, \citenamefont {Stellacci},
  \citenamefont {Genet}, \citenamefont {Schachenmayer}, \citenamefont {Genes},
  \citenamefont {Pupillo}, \citenamefont {Samor{\`\i}},\ and\ \citenamefont
  {Ebbesen}}]{orgiu2015}%
  \BibitemOpen
  \bibfield  {author} {\bibinfo {author} {\bibfnamefont {E.}~\bibnamefont
  {Orgiu}}, \bibinfo {author} {\bibfnamefont {J.}~\bibnamefont {George}},
  \bibinfo {author} {\bibfnamefont {J.~A.}\ \bibnamefont {Hutchison}}, \bibinfo
  {author} {\bibfnamefont {E.}~\bibnamefont {Devaux}}, \bibinfo {author}
  {\bibfnamefont {J.~F.}\ \bibnamefont {Dayen}}, \bibinfo {author}
  {\bibfnamefont {B.}~\bibnamefont {Doudin}}, \bibinfo {author} {\bibfnamefont
  {F.}~\bibnamefont {Stellacci}}, \bibinfo {author} {\bibfnamefont
  {C.}~\bibnamefont {Genet}}, \bibinfo {author} {\bibfnamefont
  {J.}~\bibnamefont {Schachenmayer}}, \bibinfo {author} {\bibfnamefont
  {C.}~\bibnamefont {Genes}}, \bibinfo {author} {\bibfnamefont
  {G.}~\bibnamefont {Pupillo}}, \bibinfo {author} {\bibfnamefont
  {P.}~\bibnamefont {Samor{\`\i}}}, \ and\ \bibinfo {author} {\bibfnamefont
  {T.~W.}\ \bibnamefont {Ebbesen}},\ }\href {\doibase 10.1038/nmat4392}
  {\bibfield  {journal} {\bibinfo  {journal} {Nature Materials}\ }\textbf
  {\bibinfo {volume} {14}},\ \bibinfo {pages} {1123} (\bibinfo {year}
  {2015})}\BibitemShut {NoStop}%
\bibitem [{\citenamefont {Joshi}\ \emph {et~al.}(2014)\citenamefont {Joshi},
  \citenamefont {\"Ohberg}, \citenamefont {Cresser},\ and\ \citenamefont
  {Andersson}}]{Joshi}%
  \BibitemOpen
  \bibfield  {author} {\bibinfo {author} {\bibfnamefont {C.}~\bibnamefont
  {Joshi}}, \bibinfo {author} {\bibfnamefont {P.}~\bibnamefont {\"Ohberg}},
  \bibinfo {author} {\bibfnamefont {J.~D.}\ \bibnamefont {Cresser}}, \ and\
  \bibinfo {author} {\bibfnamefont {E.}~\bibnamefont {Andersson}},\ }\href
  {\doibase 10.1103/PhysRevA.90.063815} {\bibfield  {journal} {\bibinfo
  {journal} {Phys. Rev. A}\ }\textbf {\bibinfo {volume} {90}},\ \bibinfo
  {pages} {063815} (\bibinfo {year} {2014})}\BibitemShut {NoStop}%
\bibitem [{\citenamefont {Cattaneo}\ \emph {et~al.}(2019)\citenamefont
  {Cattaneo}, \citenamefont {Giorgi}, \citenamefont {Maniscalco},\ and\
  \citenamefont {Zambrini}}]{cattaneo2019local}%
  \BibitemOpen
  \bibfield  {author} {\bibinfo {author} {\bibfnamefont {M.}~\bibnamefont
  {Cattaneo}}, \bibinfo {author} {\bibfnamefont {G.~L.}\ \bibnamefont
  {Giorgi}}, \bibinfo {author} {\bibfnamefont {S.}~\bibnamefont {Maniscalco}},
  \ and\ \bibinfo {author} {\bibfnamefont {R.}~\bibnamefont {Zambrini}},\
  }\href {\doibase 10.1088/1367-2630/ab54ac} {\bibfield  {journal} {\bibinfo
  {journal} {New Journal of Physics}\ }\textbf {\bibinfo {volume} {21}},\
  \bibinfo {pages} {113045} (\bibinfo {year} {2019})}\BibitemShut {NoStop}%
\bibitem [{\citenamefont {Levy}\ and\ \citenamefont
  {Kosloff}(2014)}]{levy2014local}%
  \BibitemOpen
  \bibfield  {author} {\bibinfo {author} {\bibfnamefont {A.}~\bibnamefont
  {Levy}}\ and\ \bibinfo {author} {\bibfnamefont {R.}~\bibnamefont {Kosloff}},\
  }\href {\doibase 10.1209/0295-5075/107/20004} {\bibfield  {journal} {\bibinfo
   {journal} {Europhysics Letters}\ }\textbf {\bibinfo {volume} {107}},\
  \bibinfo {pages} {20004} (\bibinfo {year} {2014})}\BibitemShut {NoStop}%
\bibitem [{\citenamefont {Cresser}(1992)}]{cresser1992thermal}%
  \BibitemOpen
  \bibfield  {author} {\bibinfo {author} {\bibfnamefont {J.~D.}\ \bibnamefont
  {Cresser}},\ }\href {\doibase 10.1080/09500349214552211} {\bibfield
  {journal} {\bibinfo  {journal} {Journal of Modern Optics}\ }\textbf {\bibinfo
  {volume} {39}},\ \bibinfo {pages} {2187} (\bibinfo {year}
  {1992})}\BibitemShut {NoStop}%
\bibitem [{\citenamefont {Hartmann}\ and\ \citenamefont
  {Strunz}(2020)}]{hartmann2020accuracy}%
  \BibitemOpen
  \bibfield  {author} {\bibinfo {author} {\bibfnamefont {R.}~\bibnamefont
  {Hartmann}}\ and\ \bibinfo {author} {\bibfnamefont {W.~T.}\ \bibnamefont
  {Strunz}},\ }\href {\doibase 10.1103/PhysRevA.101.012103} {\bibfield
  {journal} {\bibinfo  {journal} {Phys. Rev. A}\ }\textbf {\bibinfo {volume}
  {101}},\ \bibinfo {pages} {012103} (\bibinfo {year} {2020})}\BibitemShut
  {NoStop}%
\bibitem [{\citenamefont {Beaudoin}\ \emph {et~al.}(2011)\citenamefont
  {Beaudoin}, \citenamefont {Gambetta},\ and\ \citenamefont
  {Blais}}]{Beaudoin2011}%
  \BibitemOpen
  \bibfield  {author} {\bibinfo {author} {\bibfnamefont {F.}~\bibnamefont
  {Beaudoin}}, \bibinfo {author} {\bibfnamefont {J.~M.}\ \bibnamefont
  {Gambetta}}, \ and\ \bibinfo {author} {\bibfnamefont {A.}~\bibnamefont
  {Blais}},\ }\href {\doibase 10.1103/PhysRevA.84.043832} {\bibfield  {journal}
  {\bibinfo  {journal} {Phys. Rev. A}\ }\textbf {\bibinfo {volume} {84}},\
  \bibinfo {pages} {043832} (\bibinfo {year} {2011})}\BibitemShut {NoStop}%
\bibitem [{\citenamefont {Baust}\ \emph {et~al.}(2016)\citenamefont {Baust},
  \citenamefont {Hoffmann}, \citenamefont {Haeberlein}, \citenamefont
  {Schwarz}, \citenamefont {Eder}, \citenamefont {Goetz}, \citenamefont
  {Wulschner}, \citenamefont {Xie}, \citenamefont {Zhong}, \citenamefont
  {Quijandr\'{\i}a}, \citenamefont {Zueco}, \citenamefont {Ripoll},
  \citenamefont {Garc\'{\i}a-\'Alvarez}, \citenamefont {Romero}, \citenamefont
  {Solano}, \citenamefont {Fedorov}, \citenamefont {Menzel}, \citenamefont
  {Deppe}, \citenamefont {Marx},\ and\ \citenamefont
  {Gross}}]{baust2016ultrastrong}%
  \BibitemOpen
  \bibfield  {author} {\bibinfo {author} {\bibfnamefont {A.}~\bibnamefont
  {Baust}}, \bibinfo {author} {\bibfnamefont {E.}~\bibnamefont {Hoffmann}},
  \bibinfo {author} {\bibfnamefont {M.}~\bibnamefont {Haeberlein}}, \bibinfo
  {author} {\bibfnamefont {M.~J.}\ \bibnamefont {Schwarz}}, \bibinfo {author}
  {\bibfnamefont {P.}~\bibnamefont {Eder}}, \bibinfo {author} {\bibfnamefont
  {J.}~\bibnamefont {Goetz}}, \bibinfo {author} {\bibfnamefont
  {F.}~\bibnamefont {Wulschner}}, \bibinfo {author} {\bibfnamefont
  {E.}~\bibnamefont {Xie}}, \bibinfo {author} {\bibfnamefont {L.}~\bibnamefont
  {Zhong}}, \bibinfo {author} {\bibfnamefont {F.}~\bibnamefont
  {Quijandr\'{\i}a}}, \bibinfo {author} {\bibfnamefont {D.}~\bibnamefont
  {Zueco}}, \bibinfo {author} {\bibfnamefont {J.-J.~G.}\ \bibnamefont
  {Ripoll}}, \bibinfo {author} {\bibfnamefont {L.}~\bibnamefont
  {Garc\'{\i}a-\'Alvarez}}, \bibinfo {author} {\bibfnamefont {G.}~\bibnamefont
  {Romero}}, \bibinfo {author} {\bibfnamefont {E.}~\bibnamefont {Solano}},
  \bibinfo {author} {\bibfnamefont {K.~G.}\ \bibnamefont {Fedorov}}, \bibinfo
  {author} {\bibfnamefont {E.~P.}\ \bibnamefont {Menzel}}, \bibinfo {author}
  {\bibfnamefont {F.}~\bibnamefont {Deppe}}, \bibinfo {author} {\bibfnamefont
  {A.}~\bibnamefont {Marx}}, \ and\ \bibinfo {author} {\bibfnamefont
  {R.}~\bibnamefont {Gross}},\ }\href {\doibase 10.1103/PhysRevB.93.214501}
  {\bibfield  {journal} {\bibinfo  {journal} {Phys. Rev. B}\ }\textbf {\bibinfo
  {volume} {93}},\ \bibinfo {pages} {214501} (\bibinfo {year}
  {2016})}\BibitemShut {NoStop}%
\bibitem [{\citenamefont {Wang}\ \emph
  {et~al.}(2020{\natexlab{a}})\citenamefont {Wang}, \citenamefont {Zhang},
  \citenamefont {Wang}, \citenamefont {Chen}, \citenamefont {Li}, \citenamefont
  {Tsai}, \citenamefont {Zhu},\ and\ \citenamefont {You}}]{wang2020photon}%
  \BibitemOpen
  \bibfield  {author} {\bibinfo {author} {\bibfnamefont {S.-P.}\ \bibnamefont
  {Wang}}, \bibinfo {author} {\bibfnamefont {G.-Q.}\ \bibnamefont {Zhang}},
  \bibinfo {author} {\bibfnamefont {Y.}~\bibnamefont {Wang}}, \bibinfo {author}
  {\bibfnamefont {Z.}~\bibnamefont {Chen}}, \bibinfo {author} {\bibfnamefont
  {T.}~\bibnamefont {Li}}, \bibinfo {author} {\bibfnamefont {J.~S.}\
  \bibnamefont {Tsai}}, \bibinfo {author} {\bibfnamefont {S.-Y.}\ \bibnamefont
  {Zhu}}, \ and\ \bibinfo {author} {\bibfnamefont {J.~Q.}\ \bibnamefont
  {You}},\ }\href {\doibase 10.1103/PhysRevApplied.13.054063} {\bibfield
  {journal} {\bibinfo  {journal} {Phys. Rev. Appl.}\ }\textbf {\bibinfo
  {volume} {13}},\ \bibinfo {pages} {054063} (\bibinfo {year}
  {2020}{\natexlab{a}})}\BibitemShut {NoStop}%
\bibitem [{\citenamefont {Wallraff}\ \emph {et~al.}(2004)\citenamefont
  {Wallraff}, \citenamefont {Schuster}, \citenamefont {Blais}, \citenamefont
  {Frunzio}, \citenamefont {Huang}, \citenamefont {Majer}, \citenamefont
  {Kumar}, \citenamefont {Girvin},\ and\ \citenamefont
  {Schoelkopf}}]{wallraff2004}%
  \BibitemOpen
  \bibfield  {author} {\bibinfo {author} {\bibfnamefont {A.}~\bibnamefont
  {Wallraff}}, \bibinfo {author} {\bibfnamefont {D.~I.}\ \bibnamefont
  {Schuster}}, \bibinfo {author} {\bibfnamefont {A.}~\bibnamefont {Blais}},
  \bibinfo {author} {\bibfnamefont {L.}~\bibnamefont {Frunzio}}, \bibinfo
  {author} {\bibfnamefont {R.~S.}\ \bibnamefont {Huang}}, \bibinfo {author}
  {\bibfnamefont {J.}~\bibnamefont {Majer}}, \bibinfo {author} {\bibfnamefont
  {S.}~\bibnamefont {Kumar}}, \bibinfo {author} {\bibfnamefont {S.~M.}\
  \bibnamefont {Girvin}}, \ and\ \bibinfo {author} {\bibfnamefont {R.~J.}\
  \bibnamefont {Schoelkopf}},\ }\href {\doibase 10.1038/nature02851} {\bibfield
   {journal} {\bibinfo  {journal} {Nature}\ }\textbf {\bibinfo {volume}
  {431}},\ \bibinfo {pages} {162} (\bibinfo {year} {2004})}\BibitemShut
  {NoStop}%
\bibitem [{\citenamefont {Mergenthaler}\ \emph {et~al.}(2017)\citenamefont
  {Mergenthaler}, \citenamefont {Liu}, \citenamefont {Le~Roy}, \citenamefont
  {Ares}, \citenamefont {Thompson}, \citenamefont {Bogani}, \citenamefont
  {Luis}, \citenamefont {Blundell}, \citenamefont {Lancaster}, \citenamefont
  {Ardavan}, \citenamefont {Briggs}, \citenamefont {Leek},\ and\ \citenamefont
  {Laird}}]{mergenthaler2017}%
  \BibitemOpen
  \bibfield  {author} {\bibinfo {author} {\bibfnamefont {M.}~\bibnamefont
  {Mergenthaler}}, \bibinfo {author} {\bibfnamefont {J.}~\bibnamefont {Liu}},
  \bibinfo {author} {\bibfnamefont {J.~J.}\ \bibnamefont {Le~Roy}}, \bibinfo
  {author} {\bibfnamefont {N.}~\bibnamefont {Ares}}, \bibinfo {author}
  {\bibfnamefont {A.~L.}\ \bibnamefont {Thompson}}, \bibinfo {author}
  {\bibfnamefont {L.}~\bibnamefont {Bogani}}, \bibinfo {author} {\bibfnamefont
  {F.}~\bibnamefont {Luis}}, \bibinfo {author} {\bibfnamefont {S.~J.}\
  \bibnamefont {Blundell}}, \bibinfo {author} {\bibfnamefont {T.}~\bibnamefont
  {Lancaster}}, \bibinfo {author} {\bibfnamefont {A.}~\bibnamefont {Ardavan}},
  \bibinfo {author} {\bibfnamefont {G.~A.~D.}\ \bibnamefont {Briggs}}, \bibinfo
  {author} {\bibfnamefont {P.~J.}\ \bibnamefont {Leek}}, \ and\ \bibinfo
  {author} {\bibfnamefont {E.~A.}\ \bibnamefont {Laird}},\ }\href {\doibase
  10.1103/PhysRevLett.119.147701} {\bibfield  {journal} {\bibinfo  {journal}
  {Phys. Rev. Lett.}\ }\textbf {\bibinfo {volume} {119}},\ \bibinfo {pages}
  {147701} (\bibinfo {year} {2017})}\BibitemShut {NoStop}%
\bibitem [{\citenamefont {del R{\'\i}o}\ \emph {et~al.}(2026)\citenamefont {del
  R{\'\i}o}, \citenamefont {Rub{\'\i}n-Osanz}, \citenamefont {Rodriguez},
  \citenamefont {Roca-Jerat}, \citenamefont {Pallar{\'e}s}, \citenamefont
  {de~Sousa}, \citenamefont {Pakulski}, \citenamefont {Palacios}, \citenamefont
  {Granados}, \citenamefont {Pinkowicz}, \citenamefont {Crivillers},
  \citenamefont {Lostao}, \citenamefont {Zueco}, \citenamefont {Gomez},\ and\
  \citenamefont {Luis}}]{delrio2026}%
  \BibitemOpen
  \bibfield  {author} {\bibinfo {author} {\bibfnamefont {C.}~\bibnamefont {del
  R{\'\i}o}}, \bibinfo {author} {\bibfnamefont {M.}~\bibnamefont
  {Rub{\'\i}n-Osanz}}, \bibinfo {author} {\bibfnamefont {D.}~\bibnamefont
  {Rodriguez}}, \bibinfo {author} {\bibfnamefont {S.}~\bibnamefont
  {Roca-Jerat}}, \bibinfo {author} {\bibfnamefont {M.~C.}\ \bibnamefont
  {Pallar{\'e}s}}, \bibinfo {author} {\bibfnamefont {J.~A.}\ \bibnamefont
  {de~Sousa}}, \bibinfo {author} {\bibfnamefont {P.}~\bibnamefont {Pakulski}},
  \bibinfo {author} {\bibfnamefont {J.~L.~G.}\ \bibnamefont {Palacios}},
  \bibinfo {author} {\bibfnamefont {D.}~\bibnamefont {Granados}}, \bibinfo
  {author} {\bibfnamefont {D.}~\bibnamefont {Pinkowicz}}, \bibinfo {author}
  {\bibfnamefont {N.}~\bibnamefont {Crivillers}}, \bibinfo {author}
  {\bibfnamefont {A.}~\bibnamefont {Lostao}}, \bibinfo {author} {\bibfnamefont
  {D.}~\bibnamefont {Zueco}}, \bibinfo {author} {\bibfnamefont
  {A.}~\bibnamefont {Gomez}}, \ and\ \bibinfo {author} {\bibfnamefont
  {F.}~\bibnamefont {Luis}},\ }\href {https://arxiv.org/abs/2602.18103}
  {\enquote {\bibinfo {title} {Polariton-polariton coherent coupling in a
  molecular spin-superconductor chip},}\ } (\bibinfo {year} {2026}),\ \Eprint
  {http://arxiv.org/abs/2602.18103} {arXiv:2602.18103 [quant-ph]} \BibitemShut
  {NoStop}%
\bibitem [{\citenamefont {Takahashi}\ \emph {et~al.}(2020)\citenamefont
  {Takahashi}, \citenamefont {Kassa}, \citenamefont {Christoforou},\ and\
  \citenamefont {Keller}}]{takahashi2020strong}%
  \BibitemOpen
  \bibfield  {author} {\bibinfo {author} {\bibfnamefont {H.}~\bibnamefont
  {Takahashi}}, \bibinfo {author} {\bibfnamefont {E.}~\bibnamefont {Kassa}},
  \bibinfo {author} {\bibfnamefont {C.}~\bibnamefont {Christoforou}}, \ and\
  \bibinfo {author} {\bibfnamefont {M.}~\bibnamefont {Keller}},\ }\href
  {\doibase 10.1103/PhysRevLett.124.013602} {\bibfield  {journal} {\bibinfo
  {journal} {Phys. Rev. Lett.}\ }\textbf {\bibinfo {volume} {124}},\ \bibinfo
  {pages} {013602} (\bibinfo {year} {2020})}\BibitemShut {NoStop}%
\bibitem [{\citenamefont {Triana}\ \emph {et~al.}(2022)\citenamefont {Triana},
  \citenamefont {Arias}, \citenamefont {Nishida}, \citenamefont {Muller},
  \citenamefont {Wilcken}, \citenamefont {Johnson}, \citenamefont {Delgado},
  \citenamefont {Raschke},\ and\ \citenamefont {Herrera}}]{triana2022}%
  \BibitemOpen
  \bibfield  {author} {\bibinfo {author} {\bibfnamefont {J.~F.}\ \bibnamefont
  {Triana}}, \bibinfo {author} {\bibfnamefont {M.}~\bibnamefont {Arias}},
  \bibinfo {author} {\bibfnamefont {J.}~\bibnamefont {Nishida}}, \bibinfo
  {author} {\bibfnamefont {E.~A.}\ \bibnamefont {Muller}}, \bibinfo {author}
  {\bibfnamefont {R.}~\bibnamefont {Wilcken}}, \bibinfo {author} {\bibfnamefont
  {S.~C.}\ \bibnamefont {Johnson}}, \bibinfo {author} {\bibfnamefont
  {A.}~\bibnamefont {Delgado}}, \bibinfo {author} {\bibfnamefont {M.~B.}\
  \bibnamefont {Raschke}}, \ and\ \bibinfo {author} {\bibfnamefont
  {F.}~\bibnamefont {Herrera}},\ }\href {\doibase 10.1063/5.0075894} {\bibfield
   {journal} {\bibinfo  {journal} {The Journal of Chemical Physics}\ }\textbf
  {\bibinfo {volume} {156}},\ \bibinfo {pages} {124110} (\bibinfo {year}
  {2022})}\BibitemShut {NoStop}%
\bibitem [{\citenamefont {Chikkaraddy}\ \emph {et~al.}(2016)\citenamefont
  {Chikkaraddy}, \citenamefont {de~Nijs}, \citenamefont {Benz}, \citenamefont
  {Barrow}, \citenamefont {Scherman}, \citenamefont {Rosta}, \citenamefont
  {Demetriadou}, \citenamefont {Fox}, \citenamefont {Hess},\ and\ \citenamefont
  {Baumberg}}]{chikkarddy2016}%
  \BibitemOpen
  \bibfield  {author} {\bibinfo {author} {\bibfnamefont {R.}~\bibnamefont
  {Chikkaraddy}}, \bibinfo {author} {\bibfnamefont {B.}~\bibnamefont
  {de~Nijs}}, \bibinfo {author} {\bibfnamefont {F.}~\bibnamefont {Benz}},
  \bibinfo {author} {\bibfnamefont {S.~J.}\ \bibnamefont {Barrow}}, \bibinfo
  {author} {\bibfnamefont {O.~A.}\ \bibnamefont {Scherman}}, \bibinfo {author}
  {\bibfnamefont {E.}~\bibnamefont {Rosta}}, \bibinfo {author} {\bibfnamefont
  {A.}~\bibnamefont {Demetriadou}}, \bibinfo {author} {\bibfnamefont
  {P.}~\bibnamefont {Fox}}, \bibinfo {author} {\bibfnamefont {O.}~\bibnamefont
  {Hess}}, \ and\ \bibinfo {author} {\bibfnamefont {J.~J.}\ \bibnamefont
  {Baumberg}},\ }\href {\doibase 10.1038/nature17974} {\bibfield  {journal}
  {\bibinfo  {journal} {Nature}\ }\textbf {\bibinfo {volume} {535}},\ \bibinfo
  {pages} {127} (\bibinfo {year} {2016})}\BibitemShut {NoStop}%
\bibitem [{\citenamefont {Dann}\ and\ \citenamefont
  {Kosloff}(2021)}]{dann2021open}%
  \BibitemOpen
  \bibfield  {author} {\bibinfo {author} {\bibfnamefont {R.}~\bibnamefont
  {Dann}}\ and\ \bibinfo {author} {\bibfnamefont {R.}~\bibnamefont {Kosloff}},\
  }\href {\doibase 10.1103/PhysRevResearch.3.023006} {\bibfield  {journal}
  {\bibinfo  {journal} {Phys. Rev. Res.}\ }\textbf {\bibinfo {volume} {3}},\
  \bibinfo {pages} {023006} (\bibinfo {year} {2021})}\BibitemShut {NoStop}%
\bibitem [{\citenamefont {Cresser}\ and\ \citenamefont
  {Facer}(2017)}]{cresser2017coarse}%
  \BibitemOpen
  \bibfield  {author} {\bibinfo {author} {\bibfnamefont {J.~D.}\ \bibnamefont
  {Cresser}}\ and\ \bibinfo {author} {\bibfnamefont {C.}~\bibnamefont
  {Facer}},\ }\href {https://arxiv.org/abs/1710.09939} {\enquote {\bibinfo
  {title} {Coarse-graining in the derivation of markovian master equations and
  its significance in quantum thermodynamics},}\ } (\bibinfo {year} {2017}),\
  \Eprint {http://arxiv.org/abs/1710.09939} {arXiv:1710.09939 [quant-ph]}
  \BibitemShut {NoStop}%
\bibitem [{\citenamefont {Tscherbul}\ and\ \citenamefont
  {Brumer}(2015)}]{tscherbul2015partial}%
  \BibitemOpen
  \bibfield  {author} {\bibinfo {author} {\bibfnamefont {T.~V.}\ \bibnamefont
  {Tscherbul}}\ and\ \bibinfo {author} {\bibfnamefont {P.}~\bibnamefont
  {Brumer}},\ }\href {\doibase 10.1063/1.4908130} {\bibfield  {journal}
  {\bibinfo  {journal} {The Journal of Chemical Physics}\ }\textbf {\bibinfo
  {volume} {142}},\ \bibinfo {pages} {104107} (\bibinfo {year}
  {2015})}\BibitemShut {NoStop}%
\bibitem [{\citenamefont {Mori}\ and\ \citenamefont
  {Miyashita}(2008)}]{mori2008dynamics}%
  \BibitemOpen
  \bibfield  {author} {\bibinfo {author} {\bibfnamefont {T.}~\bibnamefont
  {Mori}}\ and\ \bibinfo {author} {\bibfnamefont {S.}~\bibnamefont
  {Miyashita}},\ }\href {\doibase 10.1143/JPSJ.77.124005} {\bibfield  {journal}
  {\bibinfo  {journal} {Journal of the Physical Society of Japan}\ }\textbf
  {\bibinfo {volume} {77}},\ \bibinfo {pages} {124005} (\bibinfo {year}
  {2008})}\BibitemShut {NoStop}%
\bibitem [{\citenamefont {Geva}\ \emph {et~al.}(2000)\citenamefont {Geva},
  \citenamefont {Rosenman},\ and\ \citenamefont {Tannor}}]{geva2000second}%
  \BibitemOpen
  \bibfield  {author} {\bibinfo {author} {\bibfnamefont {E.}~\bibnamefont
  {Geva}}, \bibinfo {author} {\bibfnamefont {E.}~\bibnamefont {Rosenman}}, \
  and\ \bibinfo {author} {\bibfnamefont {D.}~\bibnamefont {Tannor}},\ }\href
  {\doibase 10.1063/1.481928} {\bibfield  {journal} {\bibinfo  {journal} {The
  Journal of Chemical Physics}\ }\textbf {\bibinfo {volume} {113}},\ \bibinfo
  {pages} {1380} (\bibinfo {year} {2000})}\BibitemShut {NoStop}%
\bibitem [{\citenamefont {de~Vega}\ \emph {et~al.}(2010)\citenamefont
  {de~Vega}, \citenamefont {Ba{\~n}uls},\ and\ \citenamefont
  {P{\'e}rez}}]{de2010effects}%
  \BibitemOpen
  \bibfield  {author} {\bibinfo {author} {\bibfnamefont {I.}~\bibnamefont
  {de~Vega}}, \bibinfo {author} {\bibfnamefont {M.~C.}\ \bibnamefont
  {Ba{\~n}uls}}, \ and\ \bibinfo {author} {\bibfnamefont {A.}~\bibnamefont
  {P{\'e}rez}},\ }\href {\doibase 10.1088/1367-2630/12/12/123010} {\bibfield
  {journal} {\bibinfo  {journal} {New Journal of Physics}\ }\textbf {\bibinfo
  {volume} {12}},\ \bibinfo {pages} {123010} (\bibinfo {year}
  {2010})}\BibitemShut {NoStop}%
\bibitem [{\citenamefont {Farina}\ and\ \citenamefont
  {Giovannetti}(2019)}]{farina2019open}%
  \BibitemOpen
  \bibfield  {author} {\bibinfo {author} {\bibfnamefont {D.}~\bibnamefont
  {Farina}}\ and\ \bibinfo {author} {\bibfnamefont {V.}~\bibnamefont
  {Giovannetti}},\ }\href {\doibase 10.1103/PhysRevA.100.012107} {\bibfield
  {journal} {\bibinfo  {journal} {Phys. Rev. A}\ }\textbf {\bibinfo {volume}
  {100}},\ \bibinfo {pages} {012107} (\bibinfo {year} {2019})}\BibitemShut
  {NoStop}%
\bibitem [{\citenamefont {Jeske}\ \emph {et~al.}(2015)\citenamefont {Jeske},
  \citenamefont {Ing}, \citenamefont {Plenio}, \citenamefont {Huelga},\ and\
  \citenamefont {Cole}}]{jeske2015bloch}%
  \BibitemOpen
  \bibfield  {author} {\bibinfo {author} {\bibfnamefont {J.}~\bibnamefont
  {Jeske}}, \bibinfo {author} {\bibfnamefont {D.~J.}\ \bibnamefont {Ing}},
  \bibinfo {author} {\bibfnamefont {M.~B.}\ \bibnamefont {Plenio}}, \bibinfo
  {author} {\bibfnamefont {S.~F.}\ \bibnamefont {Huelga}}, \ and\ \bibinfo
  {author} {\bibfnamefont {J.~H.}\ \bibnamefont {Cole}},\ }\href {\doibase
  10.1063/1.4907370} {\bibfield  {journal} {\bibinfo  {journal} {The Journal of
  Chemical Physics}\ }\textbf {\bibinfo {volume} {142}},\ \bibinfo {pages}
  {064104} (\bibinfo {year} {2015})}\BibitemShut {NoStop}%
\bibitem [{\citenamefont {Schaller}\ and\ \citenamefont
  {Brandes}(2008)}]{schaller2008preservation}%
  \BibitemOpen
  \bibfield  {author} {\bibinfo {author} {\bibfnamefont {G.}~\bibnamefont
  {Schaller}}\ and\ \bibinfo {author} {\bibfnamefont {T.}~\bibnamefont
  {Brandes}},\ }\href {\doibase 10.1103/PhysRevA.78.022106} {\bibfield
  {journal} {\bibinfo  {journal} {Phys. Rev. A}\ }\textbf {\bibinfo {volume}
  {78}},\ \bibinfo {pages} {022106} (\bibinfo {year} {2008})}\BibitemShut
  {NoStop}%
\bibitem [{\citenamefont {Mozgunov}\ and\ \citenamefont
  {Lidar}(2020)}]{Mozgunov2020}%
  \BibitemOpen
  \bibfield  {author} {\bibinfo {author} {\bibfnamefont {E.}~\bibnamefont
  {Mozgunov}}\ and\ \bibinfo {author} {\bibfnamefont {D.}~\bibnamefont
  {Lidar}},\ }\href {\doibase 10.22331/q-2020-02-06-227} {\bibfield  {journal}
  {\bibinfo  {journal} {{Quantum}}\ }\textbf {\bibinfo {volume} {4}},\ \bibinfo
  {pages} {227} (\bibinfo {year} {2020})}\BibitemShut {NoStop}%
\bibitem [{\citenamefont {Becker}\ \emph {et~al.}(2022)\citenamefont {Becker},
  \citenamefont {Schnell},\ and\ \citenamefont
  {Thingna}}]{becker2022canonically}%
  \BibitemOpen
  \bibfield  {author} {\bibinfo {author} {\bibfnamefont {T.}~\bibnamefont
  {Becker}}, \bibinfo {author} {\bibfnamefont {A.}~\bibnamefont {Schnell}}, \
  and\ \bibinfo {author} {\bibfnamefont {J.}~\bibnamefont {Thingna}},\ }\href
  {\doibase 10.1103/PhysRevLett.129.200403} {\bibfield  {journal} {\bibinfo
  {journal} {Phys. Rev. Lett.}\ }\textbf {\bibinfo {volume} {129}},\ \bibinfo
  {pages} {200403} (\bibinfo {year} {2022})}\BibitemShut {NoStop}%
\bibitem [{\citenamefont {Potts}\ \emph {et~al.}(2021)\citenamefont {Potts},
  \citenamefont {Kalaee},\ and\ \citenamefont
  {Wacker}}]{potts2021thermodynamically}%
  \BibitemOpen
  \bibfield  {author} {\bibinfo {author} {\bibfnamefont {P.~P.}\ \bibnamefont
  {Potts}}, \bibinfo {author} {\bibfnamefont {A.~A.~S.}\ \bibnamefont
  {Kalaee}}, \ and\ \bibinfo {author} {\bibfnamefont {A.}~\bibnamefont
  {Wacker}},\ }\href {\doibase 10.1088/1367-2630/ac3b2f} {\bibfield  {journal}
  {\bibinfo  {journal} {New Journal of Physics}\ }\textbf {\bibinfo {volume}
  {23}},\ \bibinfo {pages} {123013} (\bibinfo {year} {2021})}\BibitemShut
  {NoStop}%
\bibitem [{\citenamefont {Cresser}\ and\ \citenamefont
  {Anders}(2021)}]{cresser2021weak}%
  \BibitemOpen
  \bibfield  {author} {\bibinfo {author} {\bibfnamefont {J.~D.}\ \bibnamefont
  {Cresser}}\ and\ \bibinfo {author} {\bibfnamefont {J.}~\bibnamefont
  {Anders}},\ }\href {\doibase 10.1103/PhysRevLett.127.250601} {\bibfield
  {journal} {\bibinfo  {journal} {Phys. Rev. Lett.}\ }\textbf {\bibinfo
  {volume} {127}},\ \bibinfo {pages} {250601} (\bibinfo {year}
  {2021})}\BibitemShut {NoStop}%
\bibitem [{\citenamefont {Brenes}\ \emph {et~al.}(2024)\citenamefont {Brenes},
  \citenamefont {Min}, \citenamefont {Anto-Sztrikacs}, \citenamefont
  {Bar-Gill},\ and\ \citenamefont {Segal}}]{Brenes2024}%
  \BibitemOpen
  \bibfield  {author} {\bibinfo {author} {\bibfnamefont {M.}~\bibnamefont
  {Brenes}}, \bibinfo {author} {\bibfnamefont {B.}~\bibnamefont {Min}},
  \bibinfo {author} {\bibfnamefont {N.}~\bibnamefont {Anto-Sztrikacs}},
  \bibinfo {author} {\bibfnamefont {N.}~\bibnamefont {Bar-Gill}}, \ and\
  \bibinfo {author} {\bibfnamefont {D.}~\bibnamefont {Segal}},\ }\href
  {\doibase 10.1063/5.0207028} {\bibfield  {journal} {\bibinfo  {journal} {The
  Journal of Chemical Physics}\ }\textbf {\bibinfo {volume} {160}},\ \bibinfo
  {pages} {244106} (\bibinfo {year} {2024})}\BibitemShut {NoStop}%
\bibitem [{\citenamefont {Wang}\ \emph
  {et~al.}(2020{\natexlab{b}})\citenamefont {Wang}, \citenamefont {Zhang},
  \citenamefont {Wang}, \citenamefont {Chen}, \citenamefont {Li}, \citenamefont
  {Tsai}, \citenamefont {Zhu},\ and\ \citenamefont {You}}]{wang2020}%
  \BibitemOpen
  \bibfield  {author} {\bibinfo {author} {\bibfnamefont {S.-P.}\ \bibnamefont
  {Wang}}, \bibinfo {author} {\bibfnamefont {G.-Q.}\ \bibnamefont {Zhang}},
  \bibinfo {author} {\bibfnamefont {Y.}~\bibnamefont {Wang}}, \bibinfo {author}
  {\bibfnamefont {Z.}~\bibnamefont {Chen}}, \bibinfo {author} {\bibfnamefont
  {T.}~\bibnamefont {Li}}, \bibinfo {author} {\bibfnamefont {J.~S.}\
  \bibnamefont {Tsai}}, \bibinfo {author} {\bibfnamefont {S.-Y.}\ \bibnamefont
  {Zhu}}, \ and\ \bibinfo {author} {\bibfnamefont {J.~Q.}\ \bibnamefont
  {You}},\ }\href {\doibase 10.1103/PhysRevApplied.13.054063} {\bibfield
  {journal} {\bibinfo  {journal} {Phys. Rev. Appl.}\ }\textbf {\bibinfo
  {volume} {13}},\ \bibinfo {pages} {054063} (\bibinfo {year}
  {2020}{\natexlab{b}})}\BibitemShut {NoStop}%
\bibitem [{\citenamefont {Yu}\ and\ \citenamefont
  {Frontiera}(2023)}]{yu2023ostensible}%
  \BibitemOpen
  \bibfield  {author} {\bibinfo {author} {\bibfnamefont {Z.}~\bibnamefont
  {Yu}}\ and\ \bibinfo {author} {\bibfnamefont {R.~R.}\ \bibnamefont
  {Frontiera}},\ }\href {\doibase 10.1021/acsnano.2c08630} {\bibfield
  {journal} {\bibinfo  {journal} {ACS Nano}\ }\textbf {\bibinfo {volume}
  {17}},\ \bibinfo {pages} {4306} (\bibinfo {year} {2023})}\BibitemShut
  {NoStop}%
\bibitem [{\citenamefont {Long}\ and\ \citenamefont
  {Simpkins}(2015)}]{long2015}%
  \BibitemOpen
  \bibfield  {author} {\bibinfo {author} {\bibfnamefont {J.~P.}\ \bibnamefont
  {Long}}\ and\ \bibinfo {author} {\bibfnamefont {B.~S.}\ \bibnamefont
  {Simpkins}},\ }\href {\doibase 10.1021/ph5003347} {\bibfield  {journal}
  {\bibinfo  {journal} {ACS Photonics}\ }\textbf {\bibinfo {volume} {2}},\
  \bibinfo {pages} {130} (\bibinfo {year} {2015})}\BibitemShut {NoStop}%
\bibitem [{\citenamefont {Erwin}\ \emph {et~al.}(2019)\citenamefont {Erwin},
  \citenamefont {Smotzer},\ and\ \citenamefont {Coe}}]{erwin2019}%
  \BibitemOpen
  \bibfield  {author} {\bibinfo {author} {\bibfnamefont {J.~D.}\ \bibnamefont
  {Erwin}}, \bibinfo {author} {\bibfnamefont {M.}~\bibnamefont {Smotzer}}, \
  and\ \bibinfo {author} {\bibfnamefont {J.~V.}\ \bibnamefont {Coe}},\ }\href
  {\doibase 10.1021/acs.jpcb.8b09913} {\bibfield  {journal} {\bibinfo
  {journal} {The Journal of Physical Chemistry B}\ }\textbf {\bibinfo {volume}
  {123}},\ \bibinfo {pages} {1302} (\bibinfo {year} {2019})}\BibitemShut
  {NoStop}%
\bibitem [{\citenamefont {Xiao}(2009)}]{xiao2009theory}%
  \BibitemOpen
  \bibfield  {author} {\bibinfo {author} {\bibfnamefont {M.-w.}\ \bibnamefont
  {Xiao}},\ }\href@noop {} {\bibfield  {journal} {\bibinfo  {journal} {arXiv
  preprint arXiv:0908.0787}\ } (\bibinfo {year} {2009})}\BibitemShut {NoStop}%
\end{thebibliography}%

\end{document}